\documentclass[12pt]{article}
\usepackage{graphicx}
\usepackage{lscape}
\usepackage{epsfig}
\usepackage{citesort}
\usepackage{amssymb}
\usepackage{amsmath}
\usepackage{multirow}
\usepackage[table]{xcolor}
\usepackage{colortbl}
\definecolor{lightgray}{gray}{0.9}
\newcommand{\be}{\begin{equation}}
\newcommand{\ee}{\end{equation}}
\newcommand{\bea}{\begin{eqnarray}}
\newcommand{\eea}{\end{eqnarray}}
\newcommand{\nn}{\nonumber}

\def\R1{\varepsilon_1}
\def\E8{\varepsilon_8}

\def\ga{\gamma}

\def\lb{\Lambda_b}

\def\s1{\hat s}
\def\ds{\displaystyle}

\newcommand{\bd}{\begin{displaymath}}
\newcommand{\ed}{\end{displaymath}}

\newcommand{\f}{\frac}

\def\R1{\varepsilon_1}
\def\E8{\varepsilon_8}

\def\ga{\gamma}

\def\ds{\displaystyle}
\def\beq{\begin{equation}}
\def\eeq{\end{equation}}
\def\bea{\begin{eqnarray}}
\def\eea{\end{eqnarray}}
\def\beeq{\begin{eqnarray}}
\def\eeeq{\end{eqnarray}}

\def\vel{\left|}
\def\ver{\right|}
\def\nnb{\nonumber}
\def\ga{\left(}
\def\dr{\right)}

\def\rar{\rightarrow}
\def\nnb{\nonumber}

\def\ba{\begin{array}}
\def\ea{\end{array}}

\def\xis0{{\Xi^{*0}}}

\def\g5{\gamma_5}

\def\es{\!\!\! &=& \!\!\!}

\def\ar{&+& \!\!\!}
\def\ek{&-& \!\!\!}

\newcommand{\al}{\alpha_s}

\begin{document}
\title{
         {\Large
                 {\bf Comparative analysis of the $\Lambda_b
\rightarrow \Lambda \ell^+ \ell^-$ decay  in the SM, SUSY and RS model with 
custodial protection
                 }
         }
      }
      
\author{\vspace{1cm}\\
{\small  K. Azizi$^1$ \thanks {e-mail: kazizi@dogus.edu.tr}\,\,, 
 A. T. Olgun$^2$ \thanks
{e-mail: tugba.olgun@okan.edu.tr}\,\,, Z. Tavuko\u glu$^2$ \thanks
{e-mail: zeynep.tavukoglu@okan.edu.tr}}  \\
{\small $^1$ Department of Physics, Do\u gu\c s University,
Ac{\i}badem-Kad{\i}k\"oy, 34722 \.{I}stanbul, Turkey}\\
{\small $^2$ Vocational School Kad{\i}k\"oy Campus, Okan University,
Hasanpa\c sa-Kad{\i}k\"oy, 34722 \.{I}stanbul, Turkey}\\
}

\date{}
                                 %%%%%%%%%%%%%%%%%%%%%%%%%%%%%%%%%%%%%%%
       %%%%%%%%%%%%%%%%%%%%%%%%%%%%%%%%%%%%%%%             %%%%%%%%%%%%%%%%%%%%%%%%%%%%%%%%%%%%%%%%%%
                                 %%%%%%%%%%%%%%%%%%%%%%%%%%%%%%%%%%%%%%%

\begin{titlepage}
\maketitle
\thispagestyle{empty}
\begin{abstract}
We comparatively analyze the rare $\Lambda _b\rightarrow \Lambda \ell^+ \ell^-$   channel in standard model, supersymmetry 
and Randall-Sundrum model with custodial protection (RS$_c$). Using the parametrization of the matrix elements entering the low energy 
effective Hamiltonian in terms of form factors, we calculate the corresponding differential decay width and lepton forward-backward asymmetry 
 in these models. We compare the results obtained with the most recent data from LHCb as well as lattice QCD results 
on the considered quantities. It is obtained that the standard model, with the form factors calculated in light-cone QCD sum rules, can not reproduce some experimental data on the physical quantities under consideration
  but the supersymmetry can do it. The RS$_c$ model predictions are  roughly the same as the standard model and there are no considerable differences
   between the predictions of these two models. In the case of differential decay rate, the data in the range $4 $ GeV$^2/$c$^4\leq$   $q^2 \leq 6$  GeV$^2/$c$^4$ can not be described by
   any of the considered models.
 
\end{abstract}

~~~PACS number(s): 12.60.-i,  12.60.Jv, 13.30.-a, 13.30.Ce, 14.20.Mr
\end{titlepage}

%12.60.-i -> Models beyond the standard model
%12.60.Jv ->  Supersymmetric models 
%13.30.-a -> Decays of baryons
%13.30.Ce -> Leptonic, semileptonic, and radiative decays
%14.20.Mr -> Bottom Baryons
                                 %%%%%%%%%%%%%%%%%%%%%%%%%%%%%%%%%%%%%%%
       %%%%%%%%%%%%%%%%%%%%%%%%%%%%%%%%%%%%%%%             %%%%%%%%%%%%%%%%%%%%%%%%%%%%%%%%%%%%%%%%%%
                                 %%%%%%%%%%%%%%%%%%%%%%%%%%%%%%%%%%%%%%%

\section{Introduction}

The ATLAS and CMS Collaborations at CERN have independently reported their discovery of the Higgs boson with a mass of about $125$ GeV 
using the samples of proton-proton collision data collected  in 2011 and 2012, commonly referred to as the first LHC run \cite{ATLAS,CMS,ATLASCMS}. 
Recently, a measurement of the Higgs boson mass based on the combined data samples of the ATLAS and CMS experiments has been presented 
as $m_{H}=125.09 \pm 0.21 (stat)\pm 0.11 (syst) $ GeV in Refs \cite{ATLASCMS,Higgsnew1,Higgsnew2,Higgsnew3,Higgsnew4}. At the same time,  all LHC searches for signals of new 
physics above the TeV scale  have  given  negative  results. However, the LHC constraints on new physics effects can help theoreticians in the course of searching  for these new effects and answering the questions 
that the standard model (SM) has not answered yet. We hope that  the upcoming LHC run can bring unexpected surprises to observe 
signals of new physics in the experiment \cite{Pich}.
 
Although the SM could be valid up to some arbitrary high scale, new  scenarios  should  exist  because  we  are  lacking  a  proper 
understanding  of some important issues like origin of the matter, matter-antimatter asymmetry, dark matter and dark energy etc. \cite{Pich2}. In the  baryonic sector, 
 the loop-induced flavor changing neutral current (FCNC) decay of the 
$\Lambda _b \rightarrow \Lambda \ell^+ \ell^-$ with $\ell= e, \mu, \tau$, which is  described by the $b \rightarrow s \ell^{+} \ell^{-}$ 
transition at quark level, is one of the important rare processes that can help us in the course of indirectly searching for 
new physics effects \cite{Tandean}. Recently, the differential branching fraction of the $\Lambda _b^0 \rightarrow \Lambda \mu^+ \mu^-$ 
decay channel has been measured as a function of the square of the di-muon invariant mass ($q^2$), corresponding to an integrated 
luminosity of $3.0~fb^{-1}$ using proton-proton collision data collected by the LHCb experiment \cite{LHCb}. The 
measured result at $15 $ GeV$^2/$c$^4\leq$   $q^2 \leq 20$  GeV$^2/$c$^4$ region for  the differential branching fraction  is  dBr$ (\Lambda _b^0 \rightarrow \Lambda \mu^+ \mu^-)/dq^2 = (1.18 \;^{+\,0.09}_{-\,0.08} \pm 0.03 \pm 0.27 )
\times 10^{-7}$ GeV$^2/$c$^4$.
The LHCb Collaboration has also reported  the measurement on the  forward-backward asymmetries of this transition at the $\mu$ channel. The 
measured result at the $15 $ GeV$^2/$c$^4\leq$   $q^2 \leq 20$  GeV$^2/$c$^4$ region for the lepton forward-backward asymmetry is 
$ A_{FB}^{\mu}=-0.05 \pm 0.09 (stat)\pm 0.03 (syst)$ \cite{LHCb}.  In the literature, there are a lot of 
studies on this decay channel via different approaches (for some recent studies see for instance Refs.
\cite{Gutsche,tavukoglu,Wen,Feldmann1,Guo,Mott}).
  
In the present work, we calculate the differential decay rate and lepton forward-backward asymmetry related to the FCNC 
$\Lambda _b \rightarrow \Lambda \ell^+ \ell^-$ transition for all leptons in the SM, supersymmetry (SUSY) and Randall-Sundrum scenario with custodial 
protection (RS$_c$). We compare the results with the experimental data provided by LHCb \cite{LHCb} as well as the existing lattice QCD predictions \cite{lattice}. 
Comparison of the LHCb results with the lattice QCD predictions shows that there are some deviations of data from the SM predictions. 
Such deviations can be attributed to the new physics effects that can contribute to such loop level processes. In this connection, we 
comparatively analyze the $\Lambda _b \rightarrow \Lambda \ell^+ \ell^-$ decay channel in SM and some new physics scenarios. In the calculations,
we use the form factors calculated via the light-cone QCD sum rules in \cite{form-factors}. Hence, to get ride of any misleading, we will  use SMLCSR instead of SM referring to the results that are 
obtained via using the from factors predicted by the light-cone sum rules in \cite{form-factors} when we speak about the predictions of different models.
Note that there are many studies devoted to the calculations of the form factors defining the transition under consideration via different approaches (se for instance \cite{Chen,Feldmann}), but our aim here is to
 use those form factors that are obtained in the full theory of QCD in  \cite{form-factors}  without any approximation.

The outline of the paper is as follows. In the next section, we introduce a detailed discussion of the effective Hamiltonian responsible
 for the semileptonic $\Lambda _b \rightarrow \Lambda \ell^+ \ell^-$ decay channel and  Wilson coefficients in SM, RS$_c$ and SUSY models. 
In this section, we also present a basic introduction of the RS$_c$ scenario. In section 3, we calculate the differential decay 
rate and lepton forward-backward asymmetry at  different  scenarios and compare the predictions of different models.

                                 %%%%%%%%%%%%%%%%%%%%%%%%%%%%%%%%%%%%%%%
       %%%%%%%%%%%%%%%%%%%%%%%%%%%%%%%%%%%%%%%             %%%%%%%%%%%%%%%%%%%%%%%%%%%%%%%%%%%%%%%%%%
                                 %%%%%%%%%%%%%%%%%%%%%%%%%%%%%%%%%%%%%%%

\section{The semileptonic $\Lambda _b \rightarrow \Lambda \ell^+ \ell^-$ transition in SM, SUSY and RS$_{c}$   models}
\subsection{The effective Hamiltonian and Wilson Coefficients}
At quark level, the FCNC transition of $\Lambda _b \rightarrow \Lambda \ell^+ \ell^-$ is governed by the $b \rar s \ell^+ \ell^- $ transition  
whose effective Hamiltonian in the SM can be written as
\begin{eqnarray} \label{Heff} 
{\cal H}^{eff}_{SM} &=& {G_F \alpha_{em} V_{tb}
V_{ts}^\ast \over 2\sqrt{2} \pi} \Bigg[ C^{eff,SM}_{9}
\bar{s}\gamma_\mu (1-\gamma_5) b \, \bar{\ell} \gamma^\mu \ell +
C^{SM}_{10}  \bar{s} \gamma_\mu (1-\gamma_5) b \, \bar{\ell}
\gamma^\mu
\gamma_5 \ell \nnb \\
&-&  2 m_b C^{eff,SM}_{7} {1\over q^2} \bar{s} i \sigma_{\mu\nu} q^{\nu}
(1+\gamma_5) b \, \bar{\ell} \gamma^\mu \ell \Bigg]~, 
\end{eqnarray}
where $V_{tb}$ and $V_{ts}^\ast$ are elements of the Cabibbo-Kobayashi-Maskawa (CKM) mixing matrix, $\alpha_{em}$ is the fine structure 
constant at $Z$ mass scale, $G_{F}$ is the Fermi weak coupling constant, $q^2$ is the transferred momentum squared; and the 
$C^{eff,SM}_{9}$, $C^{SM}_{10}$ and $C^{eff,SM}_{7}$ are the Wilson coefficients representing different interactions. The explicit expressions of 
the Wilson coefficients entered to the above Hamiltonian are given in the following.
%%%%%%%%%%%%%%%%%%%%%%%%%%%%%%%%%%%%%%%%%%%%%%%%%%%%%%%%%%%%%%%%%%
The Wilson coefficient $C_9^{eff,SM}$ which is a function of $\hat s^{\prime}=\frac{q^{2}}{m_{b}^{2}}$ with $q^2$ lies in the allowed region 
$4m_{l}^{2}\leq q^{2}\leq (m_{\Lambda_b}-m_{\Lambda})^{2}$  is given by \cite{Misiak,Muenz}
\begin{eqnarray} \label{wilson-C9eff}
C_9^{eff,SM}(\hat{s}^{\prime}) & = & C_9^{NDR}\eta(\hat s^{\prime}) + h(z, \hat s^{\prime})\left( 3
C_1 + C_2 + 3 C_3 + C_4 + 3C_5 + C_6 \right) \nonumber \\
& & - \f{1}{2} h(1, \hat s^{\prime}) \left( 4 C_3 + 4 C_4 + 3C_5 + C_6 \right) \nonumber \\
& & - \f{1}{2} h(0, \hat s^{\prime}) \left( C_3 + 3 C_4 \right)
+ \f{2}{9} \left( 3 C_3 + C_4 + 3 C_5 +C_6 \right)~, 
\end{eqnarray}
where the $C_9^{NDR}$ in the naive dimensional regularization (NDR) scheme is expressed as
\begin{eqnarray} \label{C9NDR}C_9^{NDR} & = & P_0^{NDR} +
\f{Y^{SM}}{\sin^2\theta_W} -4 Z^{SM} + P_E E^{SM}~. 
\end{eqnarray} 
The last term in the right hand side is neglected due to smallness of the order of $P_E$. Here $P_0^{NDR}=2.60 \pm 0.25$, $Y^{SM}=0.98$, $Z^{SM}=0.679$ and 
$\sin^2\theta_W=0.23$ \cite{Misiak,Muenz,Buras}. 
The parameter $\eta(\hat s^{\prime})$ in Eq.\eqref{wilson-C9eff} is given as
\begin{eqnarray} \eta(\hat s^{\prime}) & = & 1 + \f{\al(\mu_b)}{\pi}\,
\omega(\hat s^{\prime})~, 
\end{eqnarray} 
with
\begin{eqnarray} \label{omega-shat}
\omega(\hat s^{\prime}) & = & - \f{2}{9} \pi^2 - \f{4}{3}\mbox{Li}_2(\hat s^{\prime}) - \f{2}{3}
\ln \hat s^{\prime} \ln(1-\hat s^{\prime}) - \f{5+4\hat s^{\prime}}{3(1+2\hat s^{\prime})}
\ln(1-\hat s^{\prime}) - \nonumber \\
& &  \f{2 \hat s^{\prime} (1+\hat s^{\prime}) (1-2\hat s^{\prime})}{3(1-\hat s^{\prime})^2
(1+2\hat s^{\prime})} \ln \hat s^{\prime} + \f{5+9\hat s^{\prime}-6\hat s^{\prime2}}
{6 (1-\hat s^{\prime}) (1+2\hat s^{\prime})}~, 
\end{eqnarray}
and
\begin{eqnarray}
\alpha_s(x)=\frac{\alpha_s(m_Z)}{1-\beta_0\frac{\alpha_s(m_Z)}{2\pi}\ln(\frac{m_Z}{x})}~.
\end{eqnarray}
Here $\alpha_s(m_Z)=0.118$ and $\beta_0=\frac{23}{3}$. The function $h(y,\hat s^{\prime})$ in Eq.\eqref{wilson-C9eff} is also 
defined by
\begin{eqnarray} \label{h-phasespace} h(y,
\hat s^{\prime}) & = & -\f{8}{9}\ln\f{m_b}{\mu_b} - \f{8}{9}\ln y +
\f{8}{27} + \f{4}{9} x \\
& & - \f{2}{9} (2+x) |1-x|^{1/2} \left\{
\begin{array}{ll}
\left( \ln\left| \f{\sqrt{1-x} + 1}{\sqrt{1-x} - 1}\right| - i\pi
\right), &
\mbox{for } x \equiv \f{4z^2}{\hat s^{\prime}} < 1 \nonumber \\
2 \arctan \f{1}{\sqrt{x-1}}, & \mbox{for } x \equiv \f {4z^2}{\hat
s^{\prime}} > 1~,
\end{array}
\right. \\
\end{eqnarray}
where 
$y=1$ or $y=z=\frac{m_c}{m_b}$ and, 
\begin{eqnarray} h(0, \hat s^{\prime})
& = & \f{8}{27} -\f{8}{9} \ln\f{m_b}{\mu_b} - \f{4}{9} \ln \hat s^{\prime}
+ \f{4}{9} i\pi~.
\end{eqnarray}
In Eq.\eqref{wilson-C9eff}, the remaining coefficients are given by \cite{Buras} 
\begin{eqnarray} \label{CJ} C_j=\sum_{i=1}^8 k_{ji}
\eta^{a_i} \qquad (j=1,...6) \vspace{0.2cm}, 
\end{eqnarray}
where the $k_{ji}$ are given as \begin{eqnarray}\frac{}{}
   \label{KJI}
\begin{array}{rrrrrrrrrl}
k_{1i} = (\!\! & 0, & 0, & \f{1}{2}, & - \f{1}{2}, &
0, & 0, & 0, & 0 & \!\!)~,  \vspace{0.1cm} \\
k_{2i} = (\!\! & 0, & 0, & \f{1}{2}, &  \f{1}{2}, &
0, & 0, & 0, & 0 & \!\!)~,  \vspace{0.1cm} \\
k_{3i} = (\!\! & 0, & 0, & - \f{1}{14}, &  \f{1}{6}, &
0.0510, & - 0.1403, & - 0.0113, & 0.0054 & \!\!)~,  \vspace{0.1cm} \\
k_{4i} = (\!\! & 0, & 0, & - \f{1}{14}, &  - \f{1}{6}, &
0.0984, & 0.1214, & 0.0156, & 0.0026 & \!\!)~,  \vspace{0.1cm} \\
k_{5i} = (\!\! & 0, & 0, & 0, &  0, &
- 0.0397, & 0.0117, & - 0.0025, & 0.0304 & \!\!) ~, \vspace{0.1cm} \\
k_{6i} = (\!\! & 0, & 0, & 0, &  0, &
0.0335, & 0.0239, & - 0.0462, & -0.0112 & \!\!)~.  \vspace{0.1cm} \\
\end{array}
\end{eqnarray}
%%%%%%%%%%%%%%%%%%%%%%%%%%%%%%%%%%%%%%%%%%%%%%%%%%%%%%%%%%%%%%%%%%%%%%%%%%%%
The explicit expression for the Wilson coefficient $C^{SM}_{10}$  is given as 
\begin{eqnarray} \label{wilson-C10} 
C^{SM}_{10}= - \frac{Y^{SM}}{\sin^2 \theta_W}~. 
\end{eqnarray}
%%%%%%%%%%%%%%%%%%%%%%%%%%%%%%%%%%%%%%%%%%%%%%%%%%%%%%%%%%%%%%%%%%%%%%%%%%%%%%%%%%%%
Finally, the Wilson coefficient $C_7^{eff,SM}$ in the leading log approximation is defined 
by \cite{Misiak,Muenz,Buras,C7eff}
\begin{eqnarray}
\label{wilson-C7eff} C_7^{eff,SM}(\mu_b) \es
\eta^{\frac{16}{23}} C_7(\mu_W)+ \frac{8}{3} \left(
\eta^{\frac{14}{23}} -\eta^{\frac{16}{23}} \right) C_8(\mu_W)+C_2 (\mu_W) 
\sum_{i=1}^8 h_i \eta^{a_i}~, \nnb\\ 
\end{eqnarray}
where
\begin{eqnarray} \eta \es
\frac{\alpha_s(\mu_W)} {\alpha_s(\mu_b)}~,
\end{eqnarray}
and 
\begin{eqnarray} \label{C7C8C2}
C_7(\mu_W)=-\frac{1}{2}
D^{\prime ~SM}_{0}(x_t)~,~~ C_8(\mu_W)=-\frac{1}{2}
E^{\prime ~SM}_{0}(x_t)~,~~C_2(\mu_W)=1~. 
\end{eqnarray} 
The functions $D^{\prime~SM}_{0}(x_t)$ and $E^{\prime~SM}_{0}(x_t)$ with $x_t=\frac{m_{t}^{2}}{m_{W}^{2}}$ are given by 
\begin{eqnarray} \label{Dprime0SM} 
D^{\prime ~SM}_{0}(x_t) \es -
\frac{(8 x_t^3+5 x_t^2-7 x_t)}{12 (1-x_t)^3}
+ \frac{x_t^2(2-3 x_t)}{2(1-x_t)^4}\ln x_t~,
\end{eqnarray} 
and
\begin{eqnarray} \label{Eprime0SM} E^{\prime ~SM}_{0}(x_t) \es - \frac{x_t(x_t^2-5 x_t-2)}
{4(1-x_t)^3} + \frac{3 x_t^2}{2 (1-x_t)^4}\ln x_t~. 
\end{eqnarray} 
The coefficients $h_i$ and $a_i$ inside the $C_7^{eff,SM}$ are also given by \cite{Misiak,Muenz}
\bea\frac{}{}
   \label{hi}
\begin{array}{rrrrrrrrrl}
h_i = (\!\! & 2.2996, & - 1.0880, & - \f{3}{7}, & - \f{1}{14}, &
-0.6494, & -0.0380, & -0.0186, & -0.0057 & \!\!)  \vspace{0.1cm}, \\
\frac{}{}
   \label{ai}
a_i = (\!\! & \f{14}{23}, & \f{16}{23}, & \f{6}{23}, & -
\f{12}{23}, &
0.4086, & -0.4230, & -0.8994, & 0.1456 & \!\!).
\end{array}
\eea

One of the most important new physics scenarios is SUSY. The different SUSY models involve the SUSY I, SUSY II, SUSY III and SUSY SO(10) scenarios according to 
the values of  $tan\beta$ and an extra parameter $\mu$ with dimension of mass \cite{Aslam,Aslam2,SUSY,Wil.coef}. In SUSY I, the Wilson 
coefficient $C^{eff}_{7}$ changes its sign, the $\mu$ takes a negative value and the contributions of the neutral Higgs bosons (NHBs) 
have been disregarded. In the SUSY II model, the value of the $tan\beta$ is large and the masses of the superparticles are relatively small. 
In SUSY III, the $tan\beta$ takes a large value and the masses of the superparticles are relatively large. In SUSY SO(10), the contributions 
of the NHBs are taken into account. The supersymmetric effective Hamiltonian in terms of the new operators coming from the NHBs 
exchanges diagrams and the corresponding Wilson coefficients is written as
\begin{eqnarray} \label{HeffSUSY}\label{heff-susy} 
{\cal H}^{eff}_{SUSY} &=& {G_F \alpha_{em} V_{tb}
V_{ts}^\ast \over 2\sqrt{2} \pi} \Bigg[ C^{eff,SUSY}_{9}
\bar{s}\gamma_\mu (1-\gamma_5) b \, \bar{\ell} \gamma^\mu \ell + 
C^{\prime~eff,SUSY}_{9} \bar{s} \gamma_\mu (1+\gamma_5) b \, \bar{\ell} \gamma^\mu \ell \nnb \\
&+& C^{SUSY}_{10}  \bar{s} \gamma_\mu (1-\gamma_5) b \, \bar{\ell} \gamma^\mu \gamma_5 \ell 
+ C^{\prime~SUSY}_{10} \bar{s} \gamma_\mu (1+\gamma_5) b \, \bar{\ell} \gamma^\mu \gamma_5 \ell \nnb\\ 
&-& 2 m_b C^{eff,SUSY}_{7} {1\over q^2} \bar{s} i \sigma_{\mu\nu} q^{\nu} (1+\gamma_5) b \, \bar{\ell} \gamma^\mu \ell -  
2 m_b C^{\prime~eff,SUSY}_{7} {1\over q^2} \bar{s} i \sigma_{\mu\nu} q^{\nu} (1-\gamma_5) b \, \bar{\ell} \gamma^\mu \ell \nnb \\
&+& C^{SUSY}_{Q_1} \bar{s}  (1+\gamma_5) b \, \bar{\ell} \ell + C^{\prime~SUSY}_{Q_1} \bar{s} (1-\gamma_5) b \, \bar{\ell} \ell  \nnb \\
&+& C^{SUSY}_{Q_2} \bar{s} (1+\gamma_5) b \, \bar{\ell} \gamma_5 \ell 
+ C^{\prime~SUSY}_{Q_2} \bar{s} (1-\gamma_5) b \, \bar{\ell} \gamma_5 \ell \Bigg]~, 
\end{eqnarray}
where $C^{eff,SUSY}_{9}$, $C^{\prime~eff,SUSY}_{9}$, $C^{SUSY}_{10}$, $C^{\prime~SUSY}_{10}$, $C^{eff,SUSY}_{7}$, 
$C^{\prime~eff,SUSY}_{7}$,  $C^{SUSY}_{Q_1}$, $C^{\prime~SUSY}_{Q_1}$, $C^{SUSY}_{Q_2}$ and $C^{\prime~SUSY}_{Q_2}$ are the 
new Wilson coefficients in the different SUSY models. The new Wilson coefficients, $C^{(\prime)SUSY}_{Q_1}$ and $C^{(\prime)SUSY}_{Q_2}$ 
come from NHBs exchanging \cite{Wil.coef}. The primed Wilson coefficients only appear in SUSY SO(10) model. 
The values of Wilson coefficients in different supersymmetric models are presented in table 1 \cite{Aslam2,SUSY,Wil.coef,Aslam3}. 
\begin{table}[ht]
\centering
\rowcolors{1}{lightgray}{white}
\begin{tabular}{ccccc}
\hline \hline
\mbox{{\small Coefficient}} &\mbox{{\small SUSY I}}&\mbox{{\small SUSY II}}&\mbox{{\small SUSY III}}
&\mbox{{\small SUSY SO(10)($A_{0}=-1000$)}}
              \\
\hline
 ${\small C^{eff,SUSY}_{7}} $   &  ${\small +0.376}$ & $ {\small +0.376 }  $   
 & $ {\small -0.376} $    &$ -0.219 $                  \\
 ${\small C^{eff,SUSY}_{9}}$    &   ${\small 4.767}$ &  $ {\small 4.767 }   $  
 & $  {\small 4.767}  $   &  ${\small 4.275}   $        \\
 ${\small C^{SUSY}_{10} }    $    &  ${\small -3.735}$ & $ {\small -3.735}   $   
 & $ {\small -3.735} $    & ${\small -4.732}  $                 \\
 ${\small C^{SUSY}_{Q_1} }    $    &   ${\small 0}     $& $ {\small 6.5 (16.5)} $ 
 & $ {\small 1.2 (4.5)}$  & $ {\small 0.106+0i(1.775+0.002i)}$  \\
 ${\small C^{SUSY}_{Q_2}  }   $    &   ${\small 0}    $& ${\small -6.5 (-16.5)}$ 
 & ${\small -1.2 (-4.5)} $& ${\small -0.107+0i(-1.797-0.002i)}$ \\
 ${\small C^{\prime~eff,SUSY }_{7}}  $   & $ {\small 0} $ & $ {\small 0} $ 
 & $ {\small 0} $ & $ {\small 0.039+0.038i}$  \\
 $ {\small C^{\prime~eff,SUSY}_{9}}  $    & $ {\small 0} $ & $ {\small 0}$ 
 & $ {\small 0} $ & $ {\small 0.011+0.072i }$ \\
 ${\small C^{\prime~SUSY}_{10} }    $    & $ {\small 0} $ & $ {\small 0} $ 
 & $ {\small 0} $ & ${\small -0.075-0.67i} $  \\
 ${\small C^{\prime~SUSY}_{Q_1} }    $    & $ {\small 0} $ & $ {\small 0} $ 
 & $ {\small 0} $ & ${\small -0.247+0.242i(-4.148+4.074i)}$ \\
 ${\small C^{\prime~SUSY}_{Q_2}   }  $    & $ {\small 0} $ & $ {\small 0} $ 
 & $ {\small 0} $ & ${\small -0.25+0.246i(-4.202+4.128i)} $ \\
\hline\hline
\end{tabular}
\caption{The Wilson coefficients in different SUSY models \cite{Aslam2,SUSY,Wil.coef,Aslam3}. The values inside the parentheses 
are for the $\tau$ lepton.}
\end{table}

The last new physics scenario which we consider in this work is the Randall-Sundrum scenario proposed to solve  the 
gauge hierarchy and the flavor problems in 1999  \cite{RS1,RS2}. It is a successful model, featuring one compact extra dimension with 
non-factorizable anti-de Sitter
 ($AdS_5$) space-time \cite{Casagrande}. This model describes the five-dimensional space-time manifold with 
coordinates (x; y) and metric
\begin{eqnarray}
ds^2&=&e^{-2 k y} \eta_{\mu \nu} dx^\mu dx^\nu - dy^2 \,\,\, , \nn \\
\eta_{\mu \nu}&=&diag(+1,-1,-1,-1) \,\,\, . \label{metric}
\end{eqnarray}
The scale parameter $k$ is defined as $k\simeq {\cal O}(M_{Planck})$. We choose it as $k=10^{19}$ GeV. The fifth 
coordinate $y$ varies in a range between two branes $0$ and $L$.  $y=0$ and $y=L$ correspond to the so-called UV 
brane and IR brane, respectively. The simplest RS model with only the SM gauge group in the bulk has many important problems 
with the electroweak precision parameters \cite{Blanke}.  In the present work, we consider the RS  model  with  an  enlarged  
custodial protection  based on $SU(3)_c \times SU(2)_L \times SU(2)_R \times U(1)_{\times} \times P_{LR}$, where $P_{LR}$ 
interchanges the two $SU(2)$ groups and is responsible for the protection of the $Z b_L \overline{b_L} $ vertex (for more 
information on the model see \cite{Blanke,RSc,Casagrande,Monika,Gemmler,Duling,Albrecht,Tanedo,Scrimieri}).

The effective Hamiltonian for the  $b \rar s \ell^+ \ell^- $ transition in the RS$_{c}$ model is given as
\begin{eqnarray} \label{HeffRSc} 
{\cal H}^{eff}_{RS_{c}} &=& {G_F \alpha_{em} V_{tb}
V_{ts}^\ast \over 2\sqrt{2} \pi} \Bigg[ C^{eff,RS_c}_{9}
\bar{s}\gamma_\mu (1-\gamma_5) b \, \bar{\ell} \gamma^\mu \ell + 
C^{\prime~eff,RS_c}_{9} \bar{s} \gamma_\mu (1+\gamma_5) b \, \bar{\ell} \gamma^\mu \ell \nnb \\
&+& C^{RS_c}_{10}  \bar{s} \gamma_\mu (1-\gamma_5) b \, \bar{\ell} \gamma^\mu \gamma_5 \ell 
+ C^{\prime~RS_c}_{10} \bar{s} \gamma_\mu (1+\gamma_5) b \, \bar{\ell} \gamma^\mu \gamma_5 \ell \nnb\\ 
&-& 2 m_b C^{eff,RS_c}_{7} {1\over q^2} \bar{s} i \sigma_{\mu\nu} q^{\nu} (1+\gamma_5) b \, \bar{\ell} \gamma^\mu \ell \nnb\\
&-& 2 m_b C^{\prime~eff,RS_c}_{7} {1\over q^2} \bar{s} i \sigma_{\mu\nu} q^{\nu} (1-\gamma_5) b \, \bar{\ell} \gamma^\mu \ell \Bigg]~, 
\end{eqnarray}
where the new Wilson coefficients are modified considering the new interactions. The new coefficients in terms of the SM 
coefficients are written as \cite{Blanke,RSc,Casagrande,Monika,Gemmler,Duling,Albrecht,Tanedo,Scrimieri}
\begin{eqnarray} \label{wilson-RSc} 
C_{i}^{(\prime)RS_c}=C_{i}^{(\prime)SM}+\Delta C_{i}^{(\prime)}~,~~~~~i=7,9,10~,
\end{eqnarray}
where
\begin{eqnarray}
 \Delta C_9&=&\Bigg[\frac{\Delta Y_s}{sin^{2}(\theta_w)}-4\Delta Z_s \Bigg]~, \nnb \\
 \Delta C_9^{\prime}&=&\Bigg[\frac{\Delta Y_s^{\prime}}{sin^{2}(\theta_w)}-4\Delta Z_s^{\prime} \Bigg]~, \nnb \\
 \Delta C_{10}&=&-\frac{\Delta Y_s}{sin^{2}(\theta_w)}~, 
\end{eqnarray}
 and
 \begin{eqnarray}
 \Delta C_{10}^{\prime}&=&-\frac{\Delta Y_s^{\prime}}{sin^{2}(\theta_w)}~,
\end{eqnarray}
with
\begin{eqnarray}
 \Delta Y_{s}&=&-\frac{1}{V_{tb}V_{ts}^{\ast}}\sum_{X}\frac{\Delta_{L}^{\ell \ell}(X)-\Delta_{R}^{\ell \ell}(X)}
 {4M_{X}^{2}g_{SM}^{2}}\Delta_{L}^{bs}(X)~, \nnb \\
 \Delta Y_{s}^{\prime}&=&-\frac{1}{V_{tb}V_{ts}^{\ast}}\sum_{X}\frac{\Delta_{L}^{\ell \ell}(X)-\Delta_{R}^{\ell \ell}(X)}
 {4M_{X}^{2}g_{SM}^{2}}\Delta_{R}^{bs}(X)~, \nnb \\
 \Delta Z_{s}&=&\frac{1}{V_{tb}V_{ts}^{\ast}}\sum_{X}\frac{\Delta_{R}^{\ell \ell}(X)}
 {8M_{X}^{2}g_{SM}^{2}sin^{2}(\theta_w)}\Delta_{L}^{bs}(X)~, 
 \end{eqnarray}
 and
 \begin{eqnarray}
 \Delta Z_{s}^{\prime}&=&\frac{1}{V_{tb}V_{ts}^{\ast}}\sum_{X}\frac{\Delta_{R}^{\ell \ell}(X)}
 {8M_{X}^{2}g_{SM}^{2}sin^{2}(\theta_w)}\Delta_{R}^{bs}(X)~.
\end{eqnarray}
In the above equations, $X=Z,Z_{H},Z^{\prime}$ and $A^{(1)}$, $g_{SM}^{2}=\frac{G_F}{\sqrt{2}} \frac{\alpha}{2 \pi sin^{2}(\theta_w)}$ 
and $\theta_w$ is the Weinberg angle. The functions inside $\Delta Y_{s}$, $\Delta Y_{s}^{\prime}$, $\Delta Z_{s}$ and 
$\Delta Z_{s}^{\prime}$ are given in \cite{Blanke,RSc,Casagrande,Monika,Gemmler,Duling,Albrecht,Tanedo,Scrimieri}.

In the case of $\Delta C_{7}^{(\prime)}$,  $\Delta C_{7}^{(\prime)}(\mu_{b})=0.429 \Delta C_{7}^{(\prime)}(M_{KK})+
0.128 \Delta C_{8}^{(\prime)}(M_{KK})$ is used where the following three contributions are included \cite{RSc}:
\begin{eqnarray}
 (\Delta C_7)_1&=&i Q_u ~r \sum_{F=u,c,t}[A+2m_F^2(A^{\prime}+B^{\prime})] 
 \Big[{\cal D}_L^{\dagger}Y^u(Y^u)^{\dagger}Y^d {\cal D}_R \Big]_{23} \nnb \\
 (\Delta C_7)_2&=& -i Q_d ~r \frac{8}{3}(g_s^{4D})^2 \sum_{F=d,s,b}[I_0 +A +B +4m_F^2(I_0^{\prime}+A^{\prime}+B^{\prime})] \nnb \\
 &&\Big[{\cal D}_L^{\dagger} {\cal R}_L Y^d {\cal R}_R {\cal D}_R \Big]_{23} \nnb \\
 (\Delta C_7)_3&=& i Q_d ~r \frac{8}{3}(g_s^{4D})^2 \sum_{F=d,s,b} m_F [I_0 +A +B] 
 \Big\{\Big[{\cal D}_L^{\dagger} {\cal R}_L {\cal R}_L Y^d {\cal D}_R \Big]_{23} \nnb \\
 &+&\frac{m_b}{m_s} \Big[{\cal D}_L^{\dagger} Y^d {\cal R}_R {\cal R}_R {\cal D}_R \Big]_{23} \Big\}
\end{eqnarray}
\begin{eqnarray}
 (\Delta C_7^{\prime})_1&=&i Q_u ~r \sum_{F=u,c,t}[A+2m_F^2(A^{\prime}+B^{\prime})] 
 \Big[{\cal D}_R^{\dagger} (Y^d)^{\dagger} Y^u (Y^u)^{\dagger} {\cal D}_L \Big]_{23} \nnb \\
 (\Delta C_7^{\prime})_2&=& -i Q_d ~r \frac{8}{3}(g_s^{4D})^2 \sum_{F=d,s,b}[I_0 +A +B 
 +4m_F^2(I_0^{\prime}+A^{\prime}+B^{\prime})] \nnb \\
 &&\Big[{\cal D}_R^{\dagger} {\cal R}_R (Y^d)^{\dagger} {\cal R}_L {\cal D}_L \Big]_{23} \nnb \\
 (\Delta C_7^{\prime})_3&=& i Q_d ~r \frac{8}{3}(g_s^{4D})^2 \sum_{F=d,s,b} m_F [I_0 +A +B] 
 \Big\{\Big[{\cal D}_R^{\dagger} {\cal R}_R {\cal R}_R (Y^d)^{\dagger} {\cal D}_L \Big]_{23} \nnb \\
 &+&\frac{m_b}{m_s} \Big[{\cal D}_R^{\dagger} (Y^d)^{\dagger} {\cal R}_L {\cal R}_L {\cal D}_L \Big]_{23} \Big\}
\end{eqnarray}
\begin{eqnarray}
 (\Delta C_8)_1&=&i r \sum_{F=u,c,t}[A+2m_F^2(A^{\prime}+B^{\prime})] 
 \Big[{\cal D}_L^{\dagger}Y^u(Y^u)^{\dagger}Y^d {\cal D}_R \Big]_{23} \nnb \\
 (\Delta C_8)_2&=& -i r \frac{9}{8}(g_s^{4D})^2 \frac{v^2}{m_b m_s} {\cal T}_3 \sum_{F=d,s,b}
 [\bar{A} +\bar{B} +2m_F^2(\bar{A}^{\prime}+\bar{B}^{\prime})] \nnb \\
 &&\Big[{\cal D}_L^{\dagger} Y^d {\cal R}_R (Y^d)^{\dagger} {\cal R}_L Y^d {\cal D}_R \Big]_{23} \nnb \\
 (\Delta C_8)_3&=& -i r \frac{9}{4} (g_s^{4D})^2 {\cal T}_3  \sum_{F=d,s,b} 
 [\bar{A} +\bar{B} +2m_F^2(\bar{A}^{\prime}+\bar{B}^{\prime})] \nnb \\
&& \Big[{\cal D}_L^{\dagger} {\cal R}_L Y^d {\cal R}_R {\cal D}_R \Big]_{23} 
\end{eqnarray}
\begin{eqnarray}
 (\Delta C_8^{\prime})_1&=&i r \sum_{F=u,c,t}[A+2m_F^2(A^{\prime}+B^{\prime})] 
 \Big[{\cal D}_R^{\dagger} (Y^d)^{\dagger} Y^u (Y^u)^{\dagger} {\cal D}_L \Big]_{23} \nnb \\
 (\Delta C_8^{\prime})_2&=& -i r \frac{9}{8}(g_s^{4D})^2 \frac{v^2}{m_b m_s} {\cal T}_3 \sum_{F=d,s,b}
 [\bar{A} +\bar{B} +2m_F^2(\bar{A}^{\prime}+\bar{B}^{\prime})] \nnb \\
 &&\Big[{\cal D}_R^{\dagger} (Y^d)^{\dagger} {\cal R}_L Y^d {\cal R}_R (Y^d)^{\dagger} {\cal D}_L \Big]_{23} \nnb \\
 (\Delta C_8^{\prime})_3&=& -i r \frac{9}{4} (g_s^{4D})^2 {\cal T}_3  \sum_{F=d,s,b} 
 [\bar{A} +\bar{B} +2m_F^2(\bar{A}^{\prime}+\bar{B}^{\prime})] \nnb \\
&& \Big[{\cal D}_R^{\dagger} {\cal R}_R (Y^d)^{\dagger} {\cal R}_L {\cal D}_L \Big]_{23} 
\end{eqnarray}
where $r=\frac{v}{\frac{G_F}{4 \pi^2} V_{tb} V_{ts}^* m_b}$ and ${\cal T}_3=\frac{1}{L} \int_0^L dy[g(y)]^2$.  For the parameters inside the above equations
and the related diagrams see \cite{Blanke,RSc,Casagrande,Monika,Gemmler,Duling,Albrecht,Tanedo,Scrimieri}.
The $Q_u$ and $Q_d$ are representing the  electric charges of the up and down type quarks, respectively. 
The functions $I_0^{(\prime)}$, $A^{(\prime)}$ and $B^{(\prime)}$ are given as
\begin{eqnarray}
 I_0(t)&=& \frac{i}{(4 \pi)^2} \frac{1}{M_{KK}^2} 
 \Bigg(- \frac{1}{t-1}+\frac{ln(t)}{(t-1)^2} \Bigg) \nnb \\
 I_0^{\prime}(t)&=& \frac{i}{(4 \pi)^2} \frac{1}{M_{KK}^4} 
 \Bigg( \frac{1+t}{2t(t-1)^2}-\frac{ln(t)}{(t-1)^3} \Bigg) \nnb \\ 
 A(t)&=&B(t)=\frac{i}{(4 \pi)^2} \frac{1}{4M_{KK}^2} 
 \Bigg(\frac{t-3}{(t-1)^2}+\frac{2ln(t)}{(t-1)^3} \Bigg) \nnb \\ 
 A^{\prime}(t)&=&2B^{\prime}(t)=\frac{i}{(4 \pi)^2} \frac{1}{M_{KK}^4} 
 \Bigg(-\frac{t^2-5t-2}{6t(t-1)^3}-\frac{ln(t)}{(t-1)^4} \Bigg) \nnb \\
 \bar{A}(t)&=&\bar{B}(t)=\frac{i}{(4 \pi)^2} \frac{1}{4M_{KK}^2} 
 \Bigg(-\frac{3t-1}{(t-1)^2}+\frac{2t^2ln(t)}{(t-1)^3} \Bigg) \nnb \\
 \bar{A}^{\prime}(t)&=&\bar{B}^{\prime}(t)=\frac{i}{(4 \pi)^2} \frac{1}{4M_{KK}^4} 
 \Bigg(\frac{5t+1}{(t-1)^3}-\frac{2t(2+t)ln(t)}{(t-1)^4} \Bigg)~,
\end{eqnarray}
with $t=m_F^2 / M_{KK}^2$ (for more information see \cite{RSc}).

Fitting the parameters to the $B \rightarrow K^{*} \mu^{+} \mu^{-}$ channel, the modifications on Wilson coefficients in RS$_c$ model 
are found as table 2 \cite{RSc}.
\begin{table}[ht]
\centering
\rowcolors{1}{lightgray}{white}
\begin{tabular}{ccccccc}
\hline \hline
 &$\Delta C_{7}$ & $\Delta C_{7}^{\prime}$ & $\Delta C_{9}$ & $\Delta C_{9}^{\prime}$ 
& $\Delta C_{10}$ & $ \Delta C_{10}^{\prime}$ 
         \\
\hline \hline
$ \mbox{Values}$ & $0.046$ & $0.05$ & $0.0023$ & $0.038$ & $0.030$ & $ 0.50$   \\
 \hline \hline
\end{tabular}
\caption{The values of modifications in  Wilson coefficients in RS$_c$ model ​​used in the analysis \cite{RSc}.}
\end{table}
\subsection{Transition amplitude and matrix elements}
The amplitude of the transition under consideration is obtained by sandwiching the corresponding effective Hamiltonian between the 
initial and final baryonic states, i.e., 
\begin{eqnarray}\label{amplitude}
{\cal M}^{ \Lambda_b \rightarrow \Lambda \ell^+ \ell^-} = \langle \Lambda(p_{\Lambda}) \mid{\cal H}^{eff}\mid 
\Lambda_b(p_{\Lambda_b})\rangle~,
\end{eqnarray}
where $p_{\Lambda_b}$ and $p_{\Lambda}$ are momenta of the initial and final baryons. 
To calculate the amplitude, we need to know the following matrix elements which are parametrized in terms of twelve form factors in full QCD:
\begin{eqnarray}\label{SMtransmatrix} \langle
\Lambda(p_{\Lambda}) \mid  \bar s \gamma_\mu (1-\gamma_5) b \mid \Lambda_b(p_{\Lambda_b})\rangle\es
\bar {u}_\Lambda(p_{\Lambda}) \Bigg[\gamma_{\mu}f_{1}(q^{2})+{i}
\sigma_{\mu\nu}q^{\nu}f_{2}(q^{2}) + q^{\mu}f_{3}(q^{2}) \nnb \\
\ek \gamma_{\mu}\gamma_5
g_{1}(q^{2})-{i}\sigma_{\mu\nu}\gamma_5q^{\nu}g_{2}(q^{2})
- q^{\mu}\gamma_5 g_{3}(q^{2})
\vphantom{\int_0^{x_2}}\Bigg] u_{\Lambda_{b}}(p_{\Lambda_b})~,\nnb \\
\end{eqnarray}
\begin{eqnarray}\label{Susytransmatrix} \langle
\Lambda(p_{\Lambda}) \mid  \bar s \gamma_\mu (1+\gamma_5) b \mid \Lambda_b(p_{\Lambda_b})\rangle\es
\bar {u}_\Lambda(p_{\Lambda}) \Bigg[\gamma_{\mu}f_{1}(q^{2})+{i}
\sigma_{\mu\nu}q^{\nu}f_{2}(q^{2}) + q^{\mu}f_{3}(q^{2}) \nnb \\
&+& \gamma_{\mu}\gamma_5
g_{1}(q^{2})+{i}\sigma_{\mu\nu}\gamma_5q^{\nu}g_{2}(q^{2})
+ q^{\mu}\gamma_5 g_{3}(q^{2})
\vphantom{\int_0^{x_2}}\Bigg] u_{\Lambda_{b}}(p_{\Lambda_b})~,\nnb \\
\end{eqnarray}
\begin{eqnarray}\label{SMtransmatrix2}
\langle \Lambda(p_{\Lambda})\mid \bar s i \sigma_{\mu\nu}q^{\nu} (1+ \gamma_5)
b \mid \Lambda_b(p_{\Lambda_b})\rangle \es\bar{u}_\Lambda(p_{\Lambda})
\Bigg[\gamma_{\mu}f_{1}^{T}(q^{2})+{i}\sigma_{\mu\nu}q^{\nu}f_{2}^{T}(q^{2})+
q^{\mu}f_{3}^{T}(q^{2}) \nnb \\
\ar \gamma_{\mu}\gamma_5
g_{1}^{T}(q^{2})+{i}\sigma_{\mu\nu}\gamma_5q^{\nu}g_{2}^{T}(q^{2})
+ q^{\mu}\gamma_5 g_{3}^{T}(q^{2})
\vphantom{\int_0^{x_2}}\Bigg] u_{\Lambda_{b}}(p_{\Lambda_b})~,\nnb \\
\end{eqnarray}
\begin{eqnarray}\label{Susytransmatrix2}
\langle \Lambda(p_{\Lambda})\mid \bar s i \sigma_{\mu\nu}q^{\nu} (1- \gamma_5)
b \mid \Lambda_b(p_{\Lambda_b})\rangle \es\bar{u}_\Lambda(p_{\Lambda})
\Bigg[\gamma_{\mu}f_{1}^{T}(q^{2})+{i}\sigma_{\mu\nu}q^{\nu}f_{2}^{T}(q^{2})+
q^{\mu}f_{3}^{T}(q^{2}) \nnb \\
\ek \gamma_{\mu}\gamma_5
g_{1}^{T}(q^{2})-{i}\sigma_{\mu\nu}\gamma_5q^{\nu}g_{2}^{T}(q^{2})
- q^{\mu}\gamma_5 g_{3}^{T}(q^{2})
\vphantom{\int_0^{x_2}}\Bigg] u_{\Lambda_{b}}(p_{\Lambda_b})~,\nnb \\
\end{eqnarray}
\begin{eqnarray}\label{SUSYtransmatrix3}\langle
\Lambda(p_{\Lambda}) \mid  \bar s (1+\gamma_5) b \mid \Lambda_b(p_{\Lambda_b})\rangle\es
{1 \over m_b} \bar {u}_\Lambda(p_{\Lambda}) \Bigg[{\not\!q}f_{1}(q^{2})+{i}
q^{\mu} \sigma_{\mu\nu}q^{\nu}f_{2}(q^{2}) + q^{2}f_{3}(q^{2}) \nnb \\
\ek {\not\!q}\gamma_5
g_{1}(q^{2})-{i}q^{\mu}\sigma_{\mu\nu}\gamma_5q^{\nu}g_{2}(q^{2})
- q^{2}\gamma_5 g_{3}(q^{2})
\vphantom{\int_0^{x_2}}\Bigg] u_{\Lambda_{b}}(p_{\Lambda_b})~,\nnb \\
\end{eqnarray}
and
\begin{eqnarray}\label{SUSYtransmatrix4}\langle
\Lambda(p_{\Lambda}) \mid  \bar s (1-\gamma_5) b \mid \Lambda_b(p_{\Lambda_b})\rangle\es
{1 \over m_b} \bar {u}_\Lambda(p_{\Lambda}) \Bigg[{\not\!q} f_{1}(q^{2})+{i}
q^{\mu} \sigma_{\mu\nu}q^{\nu}f_{2}(q^{2}) + q^{2}f_{3}(q^{2}) \nnb \\
\ar {\not\!q}\gamma_5
g_{1}(q^{2})+{i}q^{\mu} \sigma_{\mu\nu}\gamma_5q^{\nu}g_{2}(q^{2})
+ q^{2}\gamma_5 g_{3}(q^{2})
\vphantom{\int_0^{x_2}}\Bigg] u_{\Lambda_{b}}(p_{\Lambda_b})~,\nnb \\
\end{eqnarray}
where the $f^{(T)}_i$ and $g^{(T)}_i$ (i running from 1 to 3) are transition form factors 
and $u_{\Lambda_b}$ and $u_{\Lambda}$ are spinors of the $\Lambda_b$ and $\Lambda$ baryons, respectively. We will use these  form factors 
from  \cite{form-factors} that have been calculated using the light-cone QCD sum rules.

Using the above transition matrix elements in terms of form factors, we find the amplitude of the transition under consideration
 at different scenarios. In the SM, we find
\begin{eqnarray}\label{amplitude1}
{\cal M}_{SM}^{ \Lambda_b \rightarrow \Lambda \ell^+ \ell^-} &=& {G_F \alpha_{em} V_{tb}V_{ts}^\ast \over 2\sqrt{2} \pi} 
\Bigg\{\Big[{\bar u}_\Lambda ({p}_{\Lambda}) ( \gamma_{\mu}[{\cal A}_1^{SM} R + {\cal B}_1^{SM} L] + 
{i}\sigma_{\mu\nu} q^{\nu}[{\cal A}_2^{SM} R + {\cal B}_2^{SM} L] \nnb \\
&+& q^{\mu} [{\cal A}_3^{SM} R + {\cal B}_3^{SM} L]) 
u_{\Lambda_{b}}(p_{\Lambda_b}) \Big] \, (\bar{\ell} 
\gamma^\mu \ell)\nnb \\
&+& \Big[{\bar u}_\Lambda ({p}_{\Lambda})( \gamma_{\mu}[{\cal D}_1^{SM} R + {\cal E}_1^{SM} L]+ {i}\sigma_{\mu\nu} q^{\nu}[{\cal D}_2^{SM} R 
+{\cal E}_2^{SM} L] \nnb \\
&+& q^{\mu} [{\cal D}_3^{SM} R + {\cal E}_3^{SM} L]) u_{\Lambda_{b}}(p_{\Lambda_b}) \Big] \,(\bar{\ell} 
\gamma^\mu \gamma_5 \ell) \Bigg\} ~.\nnb \\
\end{eqnarray}
In the case of SUSY we get
\begin{eqnarray}\label{amplitude1}
{\cal M}_{SUSY}^{ \Lambda_b \rightarrow \Lambda \ell^+ \ell^-} &=& {G_F \alpha_{em} V_{tb}V_{ts}^\ast \over 2\sqrt{2} \pi} 
\Bigg\{\Big[{\bar u}_\Lambda ({p}_{\Lambda}) ( \gamma_{\mu}[{\cal A}_1^{SUSY} R + {\cal B}_1^{SUSY} L]+ 
{i}\sigma_{\mu\nu} q^{\nu}[{\cal A}_2^{SUSY} R 
+ {\cal B}_2^{SUSY} L] \nnb \\
&+& q^{\mu} [{\cal A}_3^{SUSY} R + {\cal B}_3^{SUSY} L]) 
u_{\Lambda_{b}}(p_{\Lambda_b}) \Big] \, (\bar{\ell} \gamma^\mu \ell)\nnb \\
&+& \Big[{\bar u}_\Lambda ({p}_{\Lambda})( \gamma_{\mu}[{\cal D}_1^{SUSY} R + {\cal E}_1^{SUSY} L]
+ {i}\sigma_{\mu\nu} q^{\nu}[{\cal D}_2^{SUSY} R 
+{\cal E}_2^{SUSY} L] \nnb \\
&+& q^{\mu} [{\cal D}_3^{SUSY} R + {\cal E}_3^{SUSY} L]) 
u_{\Lambda_{b}}(p_{\Lambda_b}) \Big] \,(\bar{\ell} \gamma^\mu \gamma_5 \ell) \nnb \\
&+& \Big[{\bar u}_\Lambda ({p}_{\Lambda})( {\not\!q} [{\cal G}_1^{SUSY} R + {\cal H}_1^{SUSY} L]
+ {i} q^{\mu} \sigma_{\mu\nu} q^{\nu}[{\cal G}_2^{SUSY} R 
+ {\cal H}_2^{SUSY} L] \nnb \\ 
&+& q^2 [{\cal G}_3^{SUSY} R + {\cal H}_3^{SUSY} L]) 
u_{\Lambda_{b}}(p_{\Lambda_b}) \Big] \, (\bar{\ell} \ell)\nnb \\
&+& \Big[{\bar u}_\Lambda ({p}_{\Lambda})( {\not\!q}[{\cal K}_1^{SUSY} R + {\cal S}_1^{SUSY} L]
+ {i} q^{\mu} \sigma_{\mu\nu}q^{\nu}[{\cal K}_2^{SUSY} R 
+ {\cal S}_2^{SUSY} L] \nnb \\ 
&+& q^2 [{\cal K}_3^{SUSY} R + {\cal S}_3^{SUSY} L]) 
u_{\Lambda_{b}}(p_{\Lambda_b}) \Big] \, (\bar{\ell} \gamma_5 \ell) \Bigg\} ~,\nnb \\
\end{eqnarray}
and for RS$_c$ we obtain
\begin{eqnarray}\label{amplitude1}
{\cal M}_{RS_c}^{ \Lambda_b \rightarrow \Lambda \ell^+ \ell^-} &=& {G_F \alpha_{em} V_{tb}V_{ts}^\ast \over 2\sqrt{2} \pi} 
\Bigg\{\Big[{\bar u}_\Lambda ({p}_{\Lambda}) ( \gamma_{\mu}[{\cal A}_1^{RS_c} R + {\cal B}_1^{RS_c} L] 
+ {i}\sigma_{\mu\nu} q^{\nu}[{\cal A}_2^{RS_c} R + {\cal B}_2^{RS_c} L] \nnb \\
&+& q^{\mu} [{\cal A}_3^{RS_c} R + {\cal B}_3^{RS_c} L]) u_{\Lambda_{b}}(p_{\Lambda_b}) \Big] \, (\bar{\ell} \gamma^\mu \ell)\nnb \\
&+& \Big[{\bar u}_\Lambda ({p}_{\Lambda})( \gamma_{\mu}[{\cal D}_1^{RS_c} R + {\cal E}_1^{RS_c} L]
+ {i}\sigma_{\mu\nu} q^{\nu}[{\cal D}_2^{RS_c} R +{\cal E}_2^{RS_c} L] \nnb \\ 
&+& q^{\mu} [{\cal D}_3^{RS_c} R + {\cal E}_3^{RS_c} L]) u_{\Lambda_{b}}(p_{\Lambda_b}) \Big] \,(\bar{\ell} \gamma^\mu \gamma_5 \ell) 
\Bigg\} ~,\nnb \\
\end{eqnarray}
where $R=(1+\gamma_5)/2$ is the right-handed and $L=(1-\gamma_5)/2$ is the left-handed projectors. 
In the above equations, the calligraphic coefficients are defined at different models as
\bea \label{coef-decay-rate} {\cal A}_{1} \es f_1 C_{9}^{eff+} - g_1 C_{9}^{eff-} - 2 m_b {1\over q^2} \Big[f_1^T C_{7}^{eff+} + 
g_1^T C_{7}^{eff-} \Big] ,~ {\cal A}_{2} = {\cal A}_1 ( 1 \rar 2 ),~ {\cal A}_{3} = {\cal A}_1 \ga 1 \rar 3 \dr ~,\nnb 
\eea
\bea
{\cal B}_{1} \es f_1 C_{9}^{eff+} + g_1 C_{9}^{eff-} - 2 m_b {1\over q^2} \Big[ f_1^T C_{7}^{eff+} - 
g_1^T C_{7}^{eff-} \Big],~ {\cal B}_{2} = {\cal B}_1 \ga 1 \rar 2 \dr, ~{\cal B}_{3} = {\cal B}_1 \ga 1 \rar 3 \dr ~,\nnb 
\eea
\bea
{\cal D}_{1} = f_1 C_{10}^{+} - g_1 C_{10}^{-},~~~~~~~~~~~~~~~~ {\cal D}_{2} =  {\cal D}_1 \ga 1 \rar 2 \dr,~~~~~~~~~ 
{\cal D}_{3} = {\cal D}_1 \ga 1 \rar 3 \dr,~~~~~~~~~~~~~~~~~~~~~~~~~~~\nnb 
\eea
\bea
{\cal E}_{1} = f_1 C_{10}^{+} + g_1 C_{10}^{-},~~~~~~~~~~~~~~~~~ {\cal E}_{2} = {\cal E}_1 \ga 1 \rar 2 \dr,~~~~~~~~~~ 
{\cal E}_{3} = {\cal E}_1 \ga 1 \rar 3 \dr,~~~~~~~~~~~~~~~~~~~~~~~~~~~\nnb  
\eea
\bea
{\cal G}_{1} = {1\over m_b}  \Big[ f_1 C_{Q_1}^{+} - g_1 C_{Q_1}^{-} \Big],~~~~~~~~~ {\cal G}_{2} = {\cal G}_1 \ga 1 \rar 2 \dr,~~~~~~~~~ 
{\cal G}_{3} = {\cal G}_1 \ga 1 \rar 3 \dr,~~~~~~~~~~~~~~~~~~~~~~~~~~~\nnb 
\eea
\bea
{\cal H}_{1} = {1\over m_b} \Big[ f_1 C_{Q_1}^{+} + g_1 C_{Q_1}^{-}  \Big],~~~~~~~~ {\cal H}_{2} = {\cal H}_1 \ga 1 \rar 2 \dr,~~~~~~~~~ 
{\cal H}_{3} =  {\cal H}_1 \ga 1 \rar 3 \dr,~~~~~~~~~~~~~~~~~~~~~~~~~~\nnb 
\eea
\bea
{\cal K}_{1} = {1\over m_b} \Big[ f_1 C_{Q_2}^{+} - g_1 C_{Q_2}^{-} \Big],~~~~~~~~~ {\cal K}_{2} =  {\cal K}_1 \ga 1 \rar 2 \dr,~~~~~~~~~ 
{\cal K}_{3} =  {\cal K}_1 \ga 1 \rar 3 \dr,~~~~~~~~~~~~~~~~~~~~~~~~~~~\nnb 
\eea
\bea
{\cal S}_{1} = {1\over m_b} \Big[ f_1 C_{Q_2}^{+} + g_1 C_{Q_2}^{-} \Big],~~~~~~~~~~ {\cal S}_{2} = {\cal S}_1 \ga 1 \rar 2 \dr,~~~~~~~~~~ 
{\cal S}_{3} = {\cal S}_1 \ga 1 \rar 3 \dr,~~~~~~~~~~~~~~~~
\nnb\\
\eea
with
\bea
C_{9}^{eff+} \es C_{9}^{eff}+C^{\prime~eff}_{9},~~~~~~~~~~~~~~~~~~~~~~~~~~~~~~~~~~~~ C_{9}^{eff-} = C_{9}^{eff}-C^{\prime~eff}_{9}~, \nnb
\eea
\bea
C_{7}^{eff+} \es C_{7}^{eff}+C^{\prime~eff}_{7},~~~~~~~~~~~~~~~~~~~~~~~~~~~~~~~~~~~~ C_{7}^{eff-} = C_{7}^{eff}-C^{\prime~eff}_{7}~, \nnb
\eea
\bea
C_{10}^{+}~~~ \es C_{10}~~+~~C^{\prime}_{10},~~~~~~~~~~~~~~~~~~~~~~~~~~~~~~~~~~~~C_{10}^{-}~~~ = C_{10}~~-~~C^{\prime}_{10}~, \nnb
\eea
\bea
C_{Q_1}^{+}~~~ \es C_{Q_1}~~+~~C^{\prime}_{Q_1},~~~~~~~~~~~~~~~~~~~~~~~~~~~~~~~~~~~~C_{Q_1}^{-}~~ = C_{Q_1}~~-~~C^{\prime}_{Q_1}~, \nnb
\eea
\bea
C_{Q_2}^{+}~~~ \es C_{Q_2}~~+~~C^{\prime}_{Q_2},~~~~~~~~~~~~~~~~~~~~~~~~~~~~~~~~~~~~C_{Q_2}^{-}~~ = C_{Q_2}~~-~~C^{\prime}_{Q_2}~. 
\nnb\\
\eea

                                 %%%%%%%%%%%%%%%%%%%%%%%%%%%%%%%%%%%%%%%
       %%%%%%%%%%%%%%%%%%%%%%%%%%%%%%%%%%%%%%%             %%%%%%%%%%%%%%%%%%%%%%%%%%%%%%%%%%%%%%%%%%
                                 %%%%%%%%%%%%%%%%%%%%%%%%%%%%%%%%%%%%%%%
\section{Physical Observables}

\subsection{The differential decay width }
In the present subsection, we would like to calculate the differential decay width for the decay channel under consideration. 
Using the decay amplitude and the transition matrix elements in terms of form factors, the supersymmetric 
differential decay rate as the most comprehensive differential decay rate among the models under consideration  is obtained as
\bea\label{DDR} \frac{d^2\Gamma_{SUSY}}{d\hat
sdz}(z,\hat s) = \frac{G_F^2\alpha^2_{em} m_{\Lambda_b}}{16384
\pi^5}| V_{tb}V_{ts}^*|^2 v \sqrt{\lambda(1,r,\hat s)} \, \Bigg[{\cal
T}_{0}^{SUSY}(\hat s)+{\cal T}_{1}^{SUSY}(\hat s) z +{\cal T}_{2}^{SUSY}(\hat s)
z^2\Bigg]~, 
\nnb\\ \label{dif-decay}
\eea
where $z=\cos\theta$ with $\theta$ being the angle between the momenta of the lepton $l^+$ and the  $\Lambda_b$ in the center 
of mass of leptons, $v=\sqrt{1-\frac{4 m_\ell^2}{q^2}}$ is the lepton velocity, $\lambda=\lambda(1,r,\hat s)=(1-r-\hat s)^2-4r\hat s$ 
is the usual triangle function, $\hat s= q^2/m^2_{\Lambda_b}$ and $r= m^2_{\Lambda}/m^2_{\Lambda_b}$. The functions 
${\cal T}_{0}^{SUSY}(\hat s)$, ${\cal T}_{1}^{SUSY}(\hat s)$ and ${\cal T}_{2}^{SUSY}(\hat s)$ are obtained  as
 \bea {\cal T}_{0}^{SUSY}(\hat s) \es 32 m_\ell^2
m_{\Lambda_b}^4 \hat s (1+r-\hat s) \Big( \vel {\cal D}_{3} \ver^2 +
\vel {\cal E}_{3} \ver^2 \Big) \nnb \\
\ar 64 m_\ell^2 m_{\Lambda_b}^3 (1-r-\hat s) \, \mbox{\rm Re} \Big[{\cal D}_{1}^\ast
{\cal E}_{3} + {\cal D}_{3}
{\cal E}_1^\ast \Big] \nnb \\
\ar 64 m_{\Lambda_b}^2 \sqrt{r} (6 m_\ell^2 - m_{\Lambda_b}^2 \hat s)
{\rm Re} \Big[{\cal D}_{1}^\ast {\cal E}_{1}\Big] \nnb\\ 
\ar 64 m_\ell^2 m_{\Lambda_b}^3 \sqrt{r} \Bigg\{ 2 m_{\Lambda_b} \hat s
{\rm Re} \Big[{\cal D}_{3}^\ast {\cal E}_{3}\Big] + (1 - r + \hat s)
{\rm Re} \Big[{\cal D}_{1}^\ast {\cal D}_{3} + {\cal E}_{1}^\ast {\cal E}_{3}\Big]\Bigg\} \nnb \\
\ar 32 m_{\Lambda_b}^2 (2 m_\ell^2 + m_{\Lambda_b}^2 \hat s) \Bigg\{ (1
- r + \hat s) m_{\Lambda_b} \sqrt{r} \,
\mbox{\rm Re} \Big[{\cal A}_{1}^\ast {\cal A}_{2} + {\cal B}_{1}^\ast {\cal B}_{2}\Big] \nnb \\
\ek m_{\Lambda_b} (1 - r - \hat s) \, \mbox{\rm Re} \Big[{\cal A}_{1}^\ast {\cal B}_{2} +
{\cal A}_{2}^\ast {\cal B}_{1}\Big] - 2 \sqrt{r} \Big( \mbox{\rm Re} \Big[{\cal A}_{1}^\ast {\cal B}_{1}\Big] +
m_{\Lambda_b}^2 \hat s \,
\mbox{\rm Re} \Big[{\cal A}_{2}^\ast {\cal B}_{2}\Big] \Big) \Bigg\} \nnb \\
\ar 8 m_{\Lambda_b}^2 \Bigg\{ 4 m_\ell^2 (1 + r - \hat s) +
m_{\Lambda_b}^2 \Big[(1-r)^2 - \hat s^2 \Big]
\Bigg\} \Big( \vel {\cal A}_{1} \ver^2 +  \vel {\cal B}_{1} \ver^2 \Big) \nnb \\
\ar 8 m_{\Lambda_b}^4 \Bigg\{ 4 m_\ell^2 \Big[ \lambda + (1 + r -
\hat s) \hat s \Big] + m_{\Lambda_b}^2 \hat s \Big[(1-r)^2 - \hat s^2 \Big]
\Bigg\} \Big( \vel {\cal A}_{2} \ver^2 +  \vel {\cal B}_{2} \ver^2 \Big) \nnb \\
\ek 8 m_{\Lambda_b}^2 \Bigg\{ 4 m_\ell^2 (1 + r - \hat s) -
m_{\Lambda_b}^2 \Big[(1-r)^2 - \hat s^2 \Big]
\Bigg\} \Big( \vel {\cal D}_{1} \ver^2 +  \vel {\cal E}_{1} \ver^2 \Big) \nnb\\
\ar 8 m_{\Lambda_b}^5 \hat s v^2 \Bigg\{ - 8 m_{\Lambda_b} \hat s \sqrt{r}\,
\mbox{\rm Re} \Big[{\cal D}_{2}^\ast {\cal E}_{2}\Big] +
4 (1 - r + \hat s) \sqrt{r} \, \mbox{\rm Re}\Big[{\cal D}_{1}^\ast {\cal D}_{2}+{\cal E}_{1}^\ast {\cal E}_{2}\Big]\nnb \\
\ek 4 (1 - r - \hat s) \, \mbox{\rm Re}\Big[{\cal D}_{1}^\ast {\cal E}_{2}+{\cal D}_{2}^\ast {\cal E}_{1}\Big] +
m_{\Lambda_b} \Big[(1-r)^2 -\hat s^2\Big] \Big( \vel {\cal D}_{2} \ver^2 + \vel
{\cal E}_{2} \ver^2 \Big) \Bigg\} \nnb \\
\ek 8 m_{\Lambda_b}^4 \Bigg\{ 4 m_\ell \Big[(1-r)^2 -\hat s(1+r) \Big]\, \mbox{\rm Re} \Big[{\cal D}_{1}^\ast {\cal K}_{1}
+{\cal E}_{1}^\ast {\cal S}_{1}\Big] \nnb \\ 
\ar (4 m_\ell^2 - m_{\Lambda_b}^2 \hat s) \Big[(1-r)^2 -\hat s(1+r) \Big]\, \Big( \vel {\cal G}_{1} \ver^2 
+ \vel {\cal H}_{1} \ver^2 \Big) \nnb \\
\ar 4 m_{\Lambda_b}^2 \sqrt{r} \hat s^2 (4 m_\ell^2 
- m_{\Lambda_b}^2 \hat s) \, \mbox{\rm Re}\Big[{\cal G}_{3}^\ast {\cal H}_{3}\Big] \Bigg\} \nnb \\
\ek 8 m_{\Lambda_b}^5 \hat s \Bigg\{2 \sqrt{r} (4 m_\ell^2 
- m_{\Lambda_b}^2 \hat s) \,(1 - r + \hat s) \, \mbox{\rm Re}\Big[{\cal G}_{1}^\ast {\cal G}_{3}+{\cal H}_{1}^\ast {\cal H}_{3}\Big] \nnb \\
\ar 4 m_\ell \sqrt{r}(1 - r + \hat s) \mbox{\rm Re}\Big[{\cal D}_{1}^\ast {\cal K}_{3}
+{\cal E}_{1}^\ast {\cal S}_{3}+{\cal D}_{3}^\ast {\cal K}_{1}+{\cal E}_{3}^\ast {\cal S}_{1}\Big] \nnb 
\eea
\bea
\ar 4 m_\ell (1 - r - \hat s) \mbox{\rm Re}\Big[{\cal D}_{1}^\ast {\cal S}_{3}+{\cal E}_{1}^\ast {\cal K}_{3}
+{\cal D}_{3}^\ast {\cal S}_{1}+{\cal E}_{3}^\ast {\cal K}_{1}\Big] \nnb \\
\ar 2 (1 - r - \hat s) (4 m_\ell^2 - m_{\Lambda_b}^2 \hat s) \, \mbox{\rm Re}\Big[{\cal G}_{1}^\ast {\cal H}_{3}
+{\cal H}_{1}^\ast {\cal G}_{3}\Big] \nnb \\
\ek m_{\Lambda_b} \Big[(1-r)^2 -\hat s(1+r) \Big] \Big( \vel {\cal K}_{1} \ver^2 +  \vel {\cal S}_{1} \ver^2 \Big) \Bigg\} \nnb \\
\ek 32 m_{\Lambda_b}^4 \sqrt{r} \hat s \Bigg\{ 2 m_\ell \mbox{\rm Re}\Big[{\cal D}_{1}^\ast {\cal S}_{1}
+{\cal E}_{1}^\ast {\cal K}_{1}\Big]+(4 m_\ell^2 - m_{\Lambda_b}^2 \hat s) \,\mbox{\rm Re}\Big[{\cal G}_{1}^\ast {\cal H}_{1}\Big] \Bigg\} \nnb \\
\ar 8 m_{\Lambda_b}^6 \hat s^2 \Bigg\{ 4 \sqrt{r} \,\mbox{\rm Re}\Big[{\cal K}_{1}^\ast {\cal S}_{1}\Big]
+2 m_{\Lambda_b} \sqrt{r} (1 - r + \hat s) \mbox{\rm Re}\Big[{\cal K}_{1}^\ast {\cal K}_{3}+{\cal S}_{1}^\ast {\cal S}_{3}\Big] \nnb \\
&+& 2 m_{\Lambda_b} (1 - r - \hat s) \mbox{\rm Re}\Big[{\cal K}_{1}^\ast {\cal S}_{3}+{\cal S}_{1}^\ast {\cal K}_{3}\Big] \nnb \\
\ek (4 m_\ell^2 - m_{\Lambda_b}^2 \hat s) (1 + r - \hat s) \Big( \vel {\cal G}_{3} \ver^2 +  \vel {\cal H}_{3} \ver^2 \Big) \nnb \\
\ek 4 m_\ell (1 + r - \hat s) \mbox{\rm Re}\Big[{\cal D}_{3}^\ast {\cal K}_{3}+{\cal E}_{3}^\ast {\cal S}_{3}\Big]
- 8 m_\ell \sqrt{r} \mbox{\rm Re}\Big[{\cal D}_{3}^\ast {\cal S}_{3}+{\cal E}_{3}^\ast {\cal K}_{3}\Big]\Bigg\} \nnb \\
\ar 8 m_{\Lambda_b}^8 \hat s^3 \Bigg\{ (1 + r - \hat s) \Big( \vel {\cal K}_{3} \ver^2 +  \vel {\cal S}_{3} \ver^2 \Big) 
+ 4 \sqrt{r} \mbox{\rm Re}\Big[{\cal K}_{3}^\ast {\cal S}_{3}\Big]\Bigg\}, \nnb \\
\eea
\bea {\cal T}_{1}^{SUSY}(\hat s) &=& -32
m_{\lb}^4 m_\ell \sqrt{\lambda} v (1 - r)
Re\Big({\cal A}_{1}^* {\cal G}_{1}+{\cal B}_{1}^* {\cal H}_{1}\Big)\nn\\
&-&16 m_{\lb}^4\s1 v \sqrt{\lambda} 
\Bigg\{ 2 Re\Big({\cal A}_{1}^* {\cal D}_{1}\Big)-2Re\Big({\cal B}_{1}^* {\cal E}_{1}\Big)\nn\\
&+&2 m_{\lb} Re\Big({\cal B}_{1}^* {\cal D}_{2}-{\cal B}_{2}^* {\cal D}_{1}+{\cal A}_{2}^* {\cal E}_{1}
-{\cal A}_{1}^*{\cal E}_{2}\Big)\nn\\
&+&2 m_{\lb} m_\ell Re\Big({\cal A}_{1}^* {\cal H}_{3}+{\cal B}_{1}^* {\cal G}_{3}-{\cal A}_{2}^* {\cal H}_{1}
-{\cal B}_{2}^*{\cal G}_{1}\Big)\Bigg\}\nn\\
&+&32 m_{\lb}^5 \s1~ v \sqrt{\lambda} \Bigg\{
m_{\lb} (1-r)Re\Big({\cal A}_{2}^* {\cal D}_{2} -{\cal B}_{2}^* {\cal E}_{2}\Big)\nn\\
&+& \sqrt{r} Re\Big({\cal A}_{2}^* {\cal D}_{1}+{\cal A}_{1}^* {\cal D}_{2}-{\cal B}_{2}^*{\cal E}_{1}
-{\cal B}_{1}^* {\cal E}_{2}\Big)\nn\\
&-& \sqrt{r} m_\ell Re\Big({\cal A}_{1}^* {\cal G}_{3}+{\cal B}_{1}^* {\cal H}_{3}+{\cal A}_{2}^*{\cal G}_{1}
+{\cal B}_{2}^* {\cal H}_{1}\Big)\Bigg\} \nn\\
&+&32 m_{\lb}^6 m_\ell \sqrt{\lambda} v \hat s^2 Re\Big({\cal A}_{2}^* {\cal G}_{3}+{\cal B}_{2}^* {\cal H}_{3}\Big),\nn\\
\eea\
and
\bea {\cal T}_{2}^{SUSY}(\hat s) \es - 8 m_{\Lambda_b}^4 v^2 \lambda \Big(\vel {\cal A}_{1} \ver^2 + \vel {\cal B}_{1} \ver^2 
+ \vel {\cal D}_{1} \ver^2 + \vel {\cal E}_{1} \ver^2 \Big) \nnb \\
\ar 8 m_{\Lambda_b}^6 \hat s v^2 \lambda \Big( \vel {\cal A}_{2} \ver^2 + \vel
{\cal B}_{2} \ver^2 + \vel {\cal D}_{2} \ver^2 + \vel {\cal E}_{2} \ver^2 \Big) ~.\nn\\ 
\eea\
Integrating the Eq.\eqref{DDR} over $z$ in the interval $[-1,1]$, we obtain the differential decay width only in terms of $ \hat s$ as 
\begin{eqnarray}
\frac{d\Gamma_{SUSY}}{d \hat s} (\hat s)= \frac{G_F^2\alpha^2_{em} m_{\Lambda_b}}{8192
\pi^5}| V_{tb}V_{ts}^*|^2 v \sqrt{\lambda} \, \Bigg[{{\cal T}_0^{SUSY}(\hat s)
+\frac{1}{3} {\cal T}_2^{SUSY}(\hat s)}\Bigg]~. \label{decayrate} 
\end{eqnarray}
The differential decay rate of RS$_c$ is found from $\frac{d\Gamma_{SUSY}}{d \hat s} (\hat s)$ by replacing $C_{Q_1}$, 
$C^{\prime}_{Q_1}$, $C_{Q_2}$ and $C^{\prime}_{Q_2}$ with zero. In the case of SM, $\frac{d\Gamma_{SM}}{d \hat s} (\hat s)$ is found from the 
supersymmetric differential decay rate via setting $C^{\prime~eff}_{7}$, $C^{\prime~eff}_{9}$, $C^{\prime}_{10}$, $C_{Q_1}$, 
$C^{\prime}_{Q_1}$, $C_{Q_2}$ and $C^{\prime}_{Q_2}$ to zero.
\subsection{The differential branching ratio }
In this subsection, we numerically analyze the differential branching ratio that depends on $q^2$ for the $\Lambda_{b}\rightarrow \Lambda \ell^{+} \ell^{-}$ decay in  SMLCSR, SUSY and RS$_c$ scenarios. In order to discuss the variation of the differential branching ratio  with respect to $q^2$, we shall present some values of input parameters  in table 3
besides the form factors as the main inputs.
\begin{table}[ht]
\centering
\rowcolors{1}{lightgray}{white}
\begin{tabular}{cc}
\hline \hline
   Some Input Parameters  &  Values    
           \\
\hline \hline
$ m_\mu $            &   $ 0.10565                         $   GeV \\
$ m_\tau $           &   $ 1.77682                         $   GeV \\
$ m_c $              &   $ 1.275 \pm 0.025                 $   GeV \\
$ m_b $              &   $ 4.18  \pm 0.03                  $   GeV \\
$ m_t $              &   $ 173.21 \pm 0.51 \pm 0.71        $   GeV \\
$ m_W $              &   $ 80.385 \pm 0.015                $   GeV \\
$ m_{\Lambda_b} $    &   $ 5.6195 \pm 0.0004               $   GeV \\
$ m_{\Lambda} $      &   $ 1.11568                         $   GeV\\
$ \tau_{\Lambda_b} $ &   $ (1.451\pm0.013)\times 10^{-12}  $   s      \\
$ \hbar  $           &   $ 6.582\times 10^{-25}            $   GeV s   \\
$ G_{F} $            &   $ 1.166\times 10^{-5}             $   GeV$^{-2}$ \\
$ \alpha_{em} $      &   $ 1/137                           $               \\
$ | V_{tb}V_{ts}^*|$ &   $ 0.040                           $                \\
 \hline \hline
 
\end{tabular}
\caption{The values of some input parameters ​​used in our calculations, taken generally from PDG \cite{PDG}.}
\end{table}
\begin{figure}[h!]
\centering
\begin{tabular}{cc}
\epsfig{file=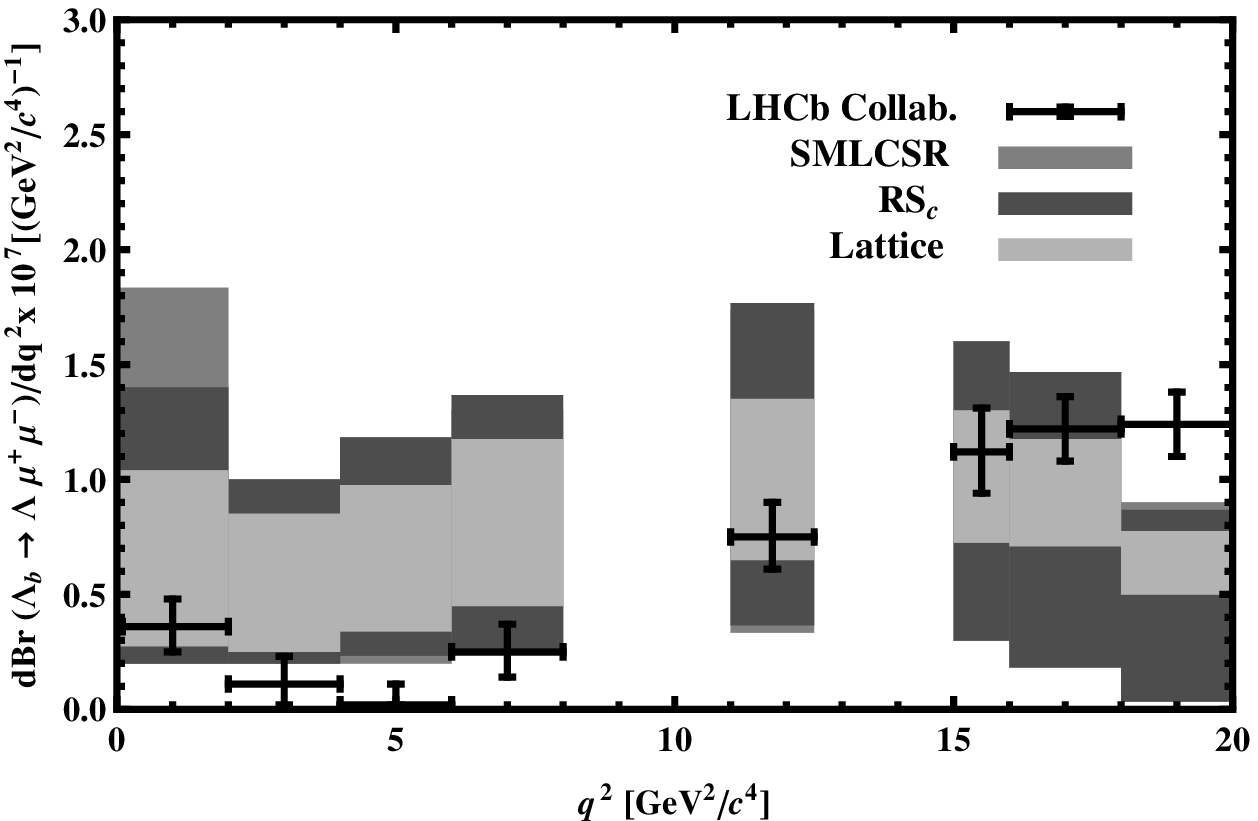,width=0.55\linewidth,clip=} & 
\end{tabular}
\caption{The dependence of the differential branching ratio on  $q^2$  for the $\Lambda_{b}\rightarrow \Lambda \mu^{+} \mu^{-}$ 
transition in the SMLCSR and RS$_c$ models. The lattice QCD \cite{lattice} and recent experimental data by LHCb \cite{LHCb} Collaboration 
are also included.}
\end{figure}
\begin{figure}[h!]
\centering
\begin{tabular}{cc}
\epsfig{file=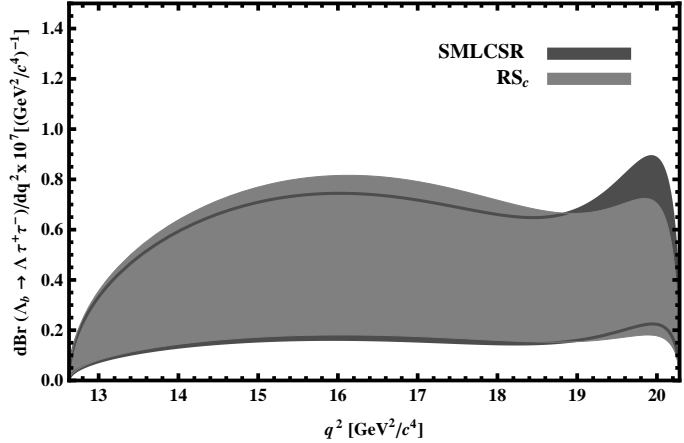,width=0.55\linewidth,clip=} & 
\end{tabular}
\caption{The dependence of the differential branching ratio on  $q^2$  for the $\Lambda_{b}\rightarrow \Lambda \tau^{+} \tau^{-}$ 
transition in the SMLCSR and RS$_c$ models.}
\end{figure}
\begin{figure}[h!]
\centering
\begin{tabular}{ccc}
\epsfig{file=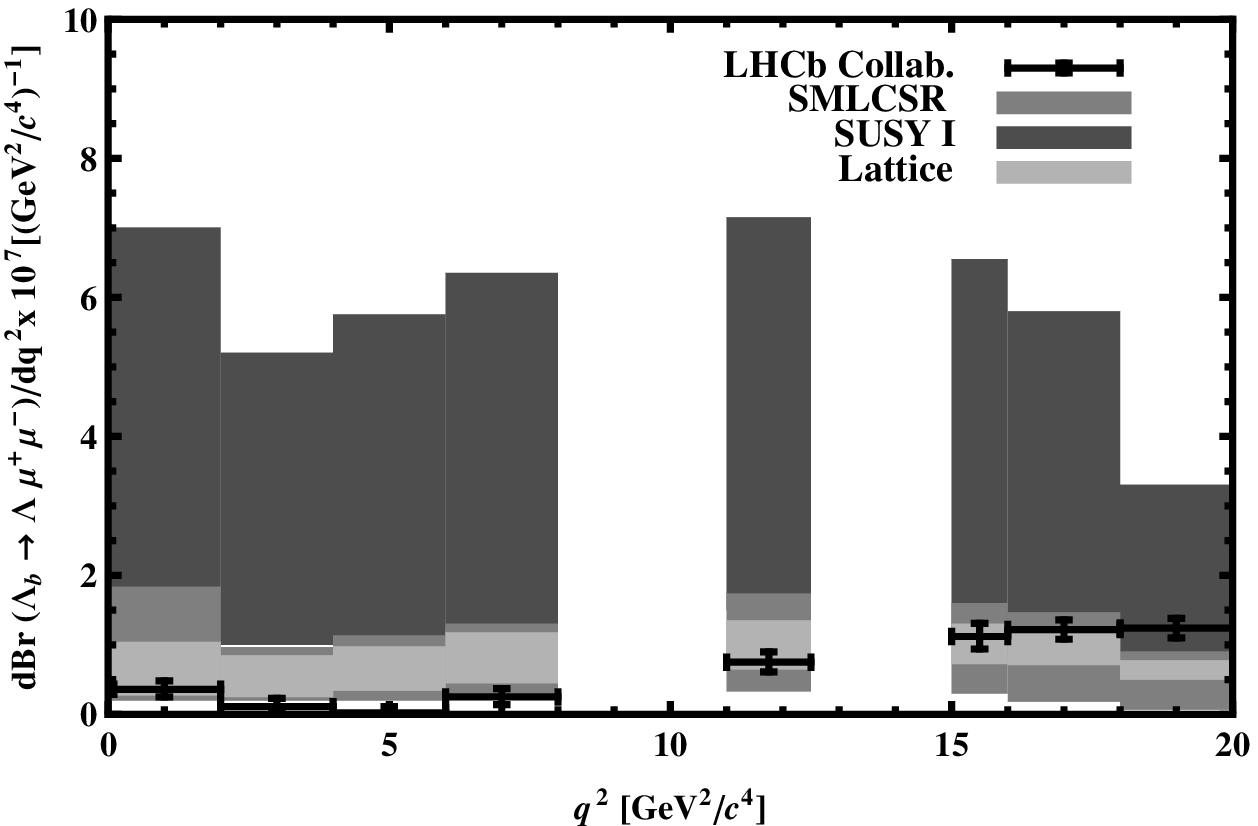,width=0.45\linewidth,clip=} &
\epsfig{file=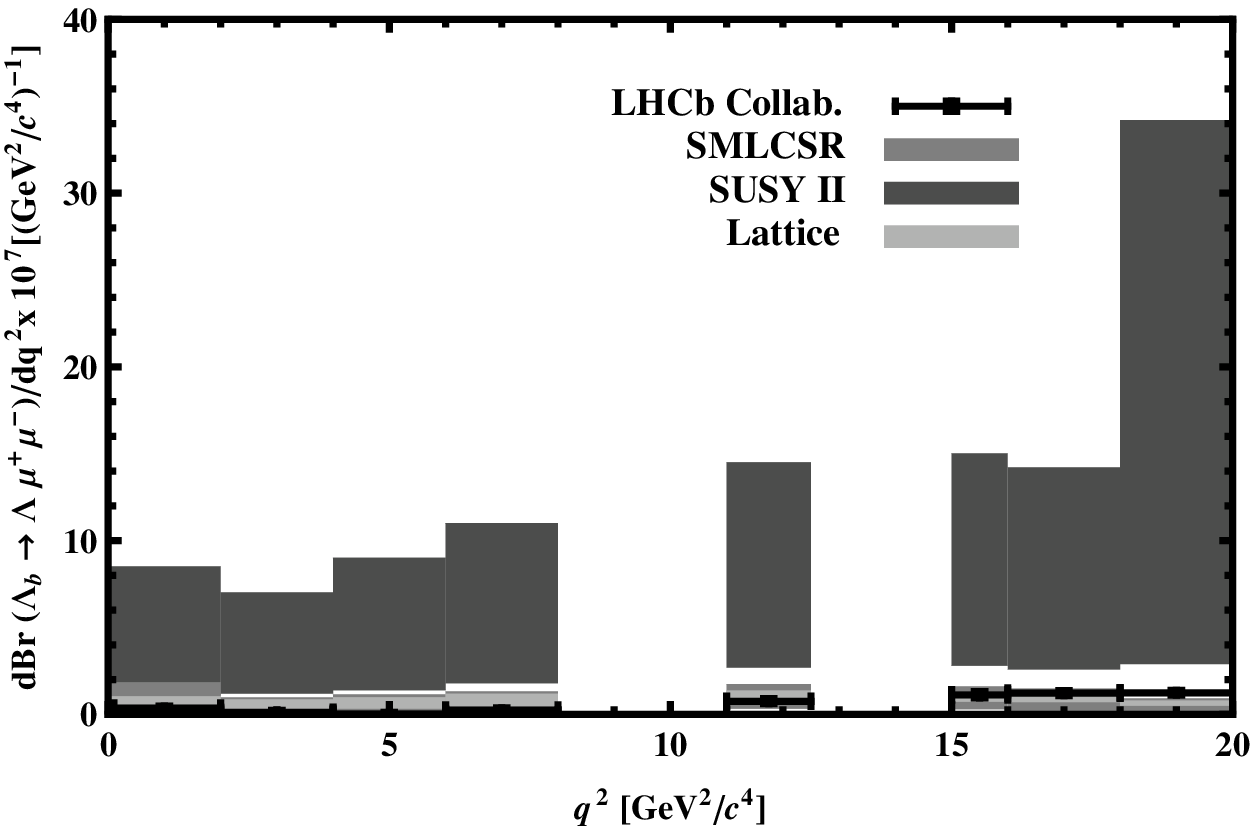,width=0.45\linewidth,clip=} 
\end{tabular}
\caption{The dependence of the differential branching ratio on  $q^2$  for the $\Lambda_{b}\rightarrow \Lambda \mu^{+} \mu^{-}$ 
transition in SMLCSR and SUSY I and II models. The lattice QCD \cite{lattice} and recent experimental data by LHCb \cite{LHCb} Collaboration 
are also included.}
\end{figure}
\begin{figure}[h!]
\centering
\begin{tabular}{ccc}
\epsfig{file=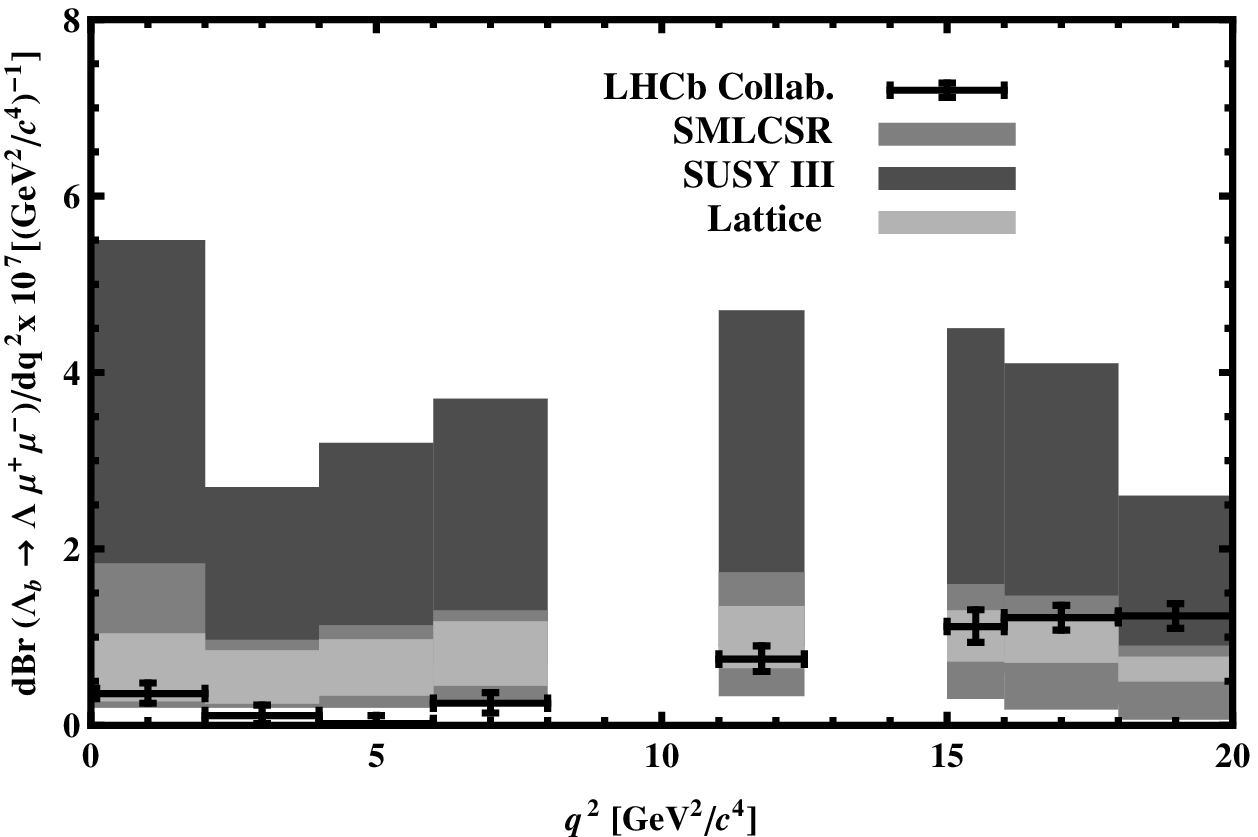,width=0.45\linewidth,clip=} &
\epsfig{file=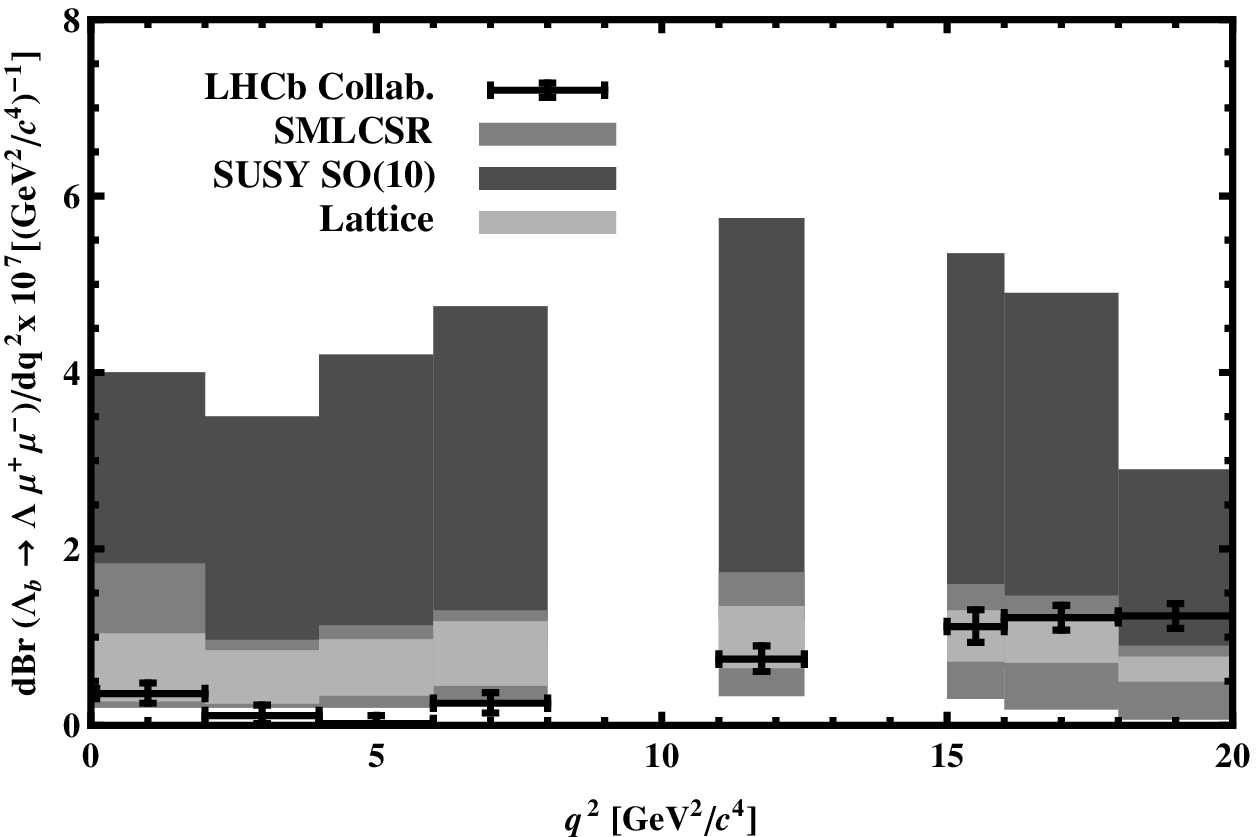,width=0.45\linewidth,clip=} 
\end{tabular}
\caption{The dependence of the differential branching ratio on  $q^2$  for the $\Lambda_{b}\rightarrow \Lambda \mu^{+} \mu^{-}$ 
transition in SMLCSR and  SUSY III and SO(10) models. The lattice QCD \cite{lattice} and recent experimental data by LHCb \cite{LHCb} Collaboration 
are also included.}
\end{figure}
\begin{figure}[h!]
\centering
\begin{tabular}{ccc}
\epsfig{file=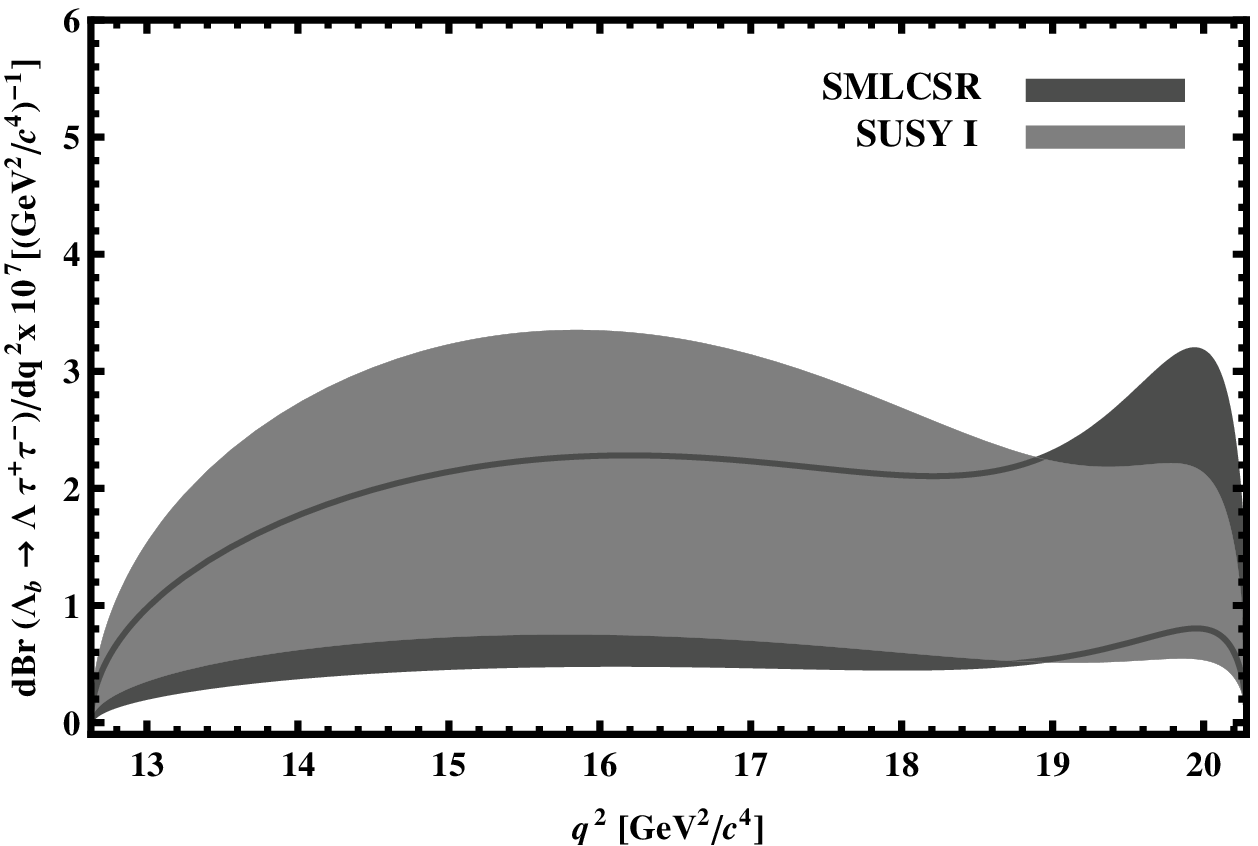,width=0.45\linewidth,clip=} &
\epsfig{file=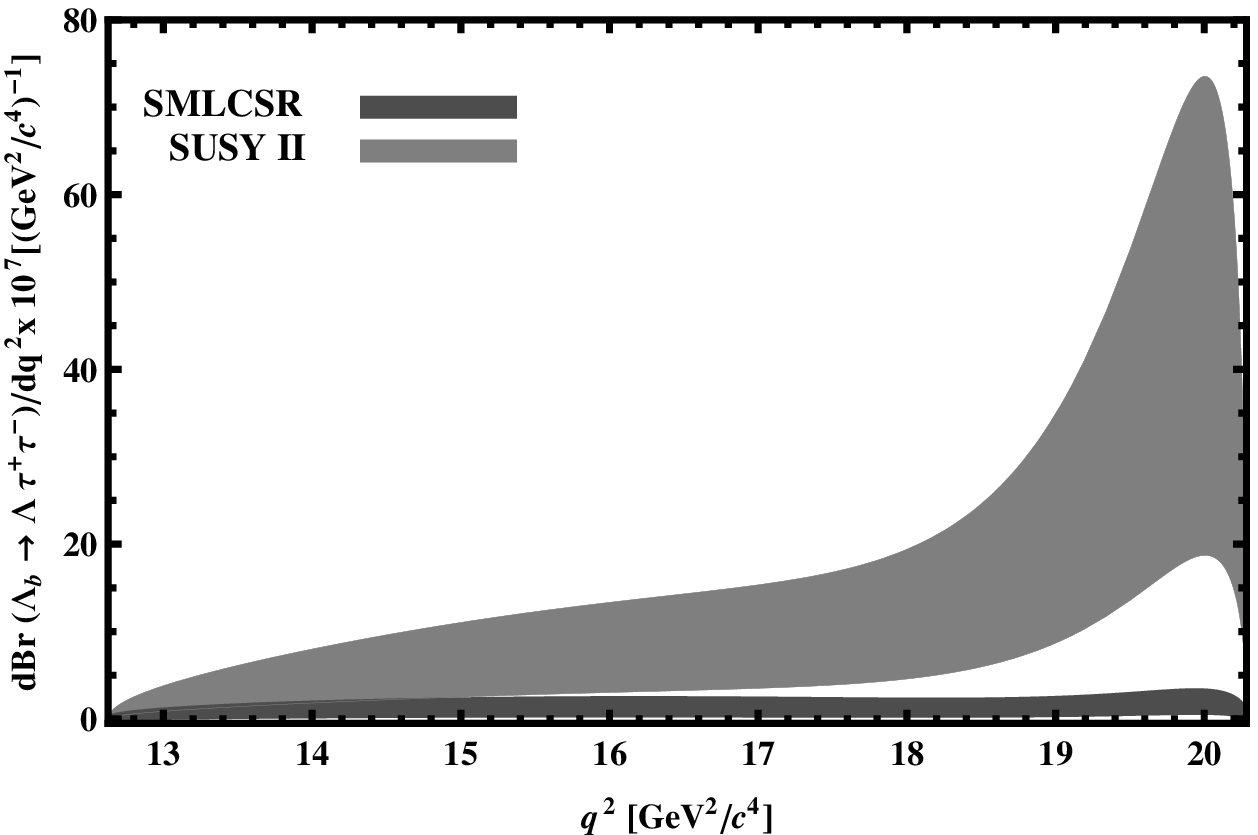,width=0.45\linewidth,clip=} 
\end{tabular}
\caption{The dependence of the differential branching ratio on  $q^2$  for the $\Lambda_{b}\rightarrow \Lambda \tau^{+} \tau^{-}$ 
transition in SMLCSR and SUSY I and II models.}
\end{figure}
\begin{figure}[h!]
\centering
\begin{tabular}{ccc}
\epsfig{file=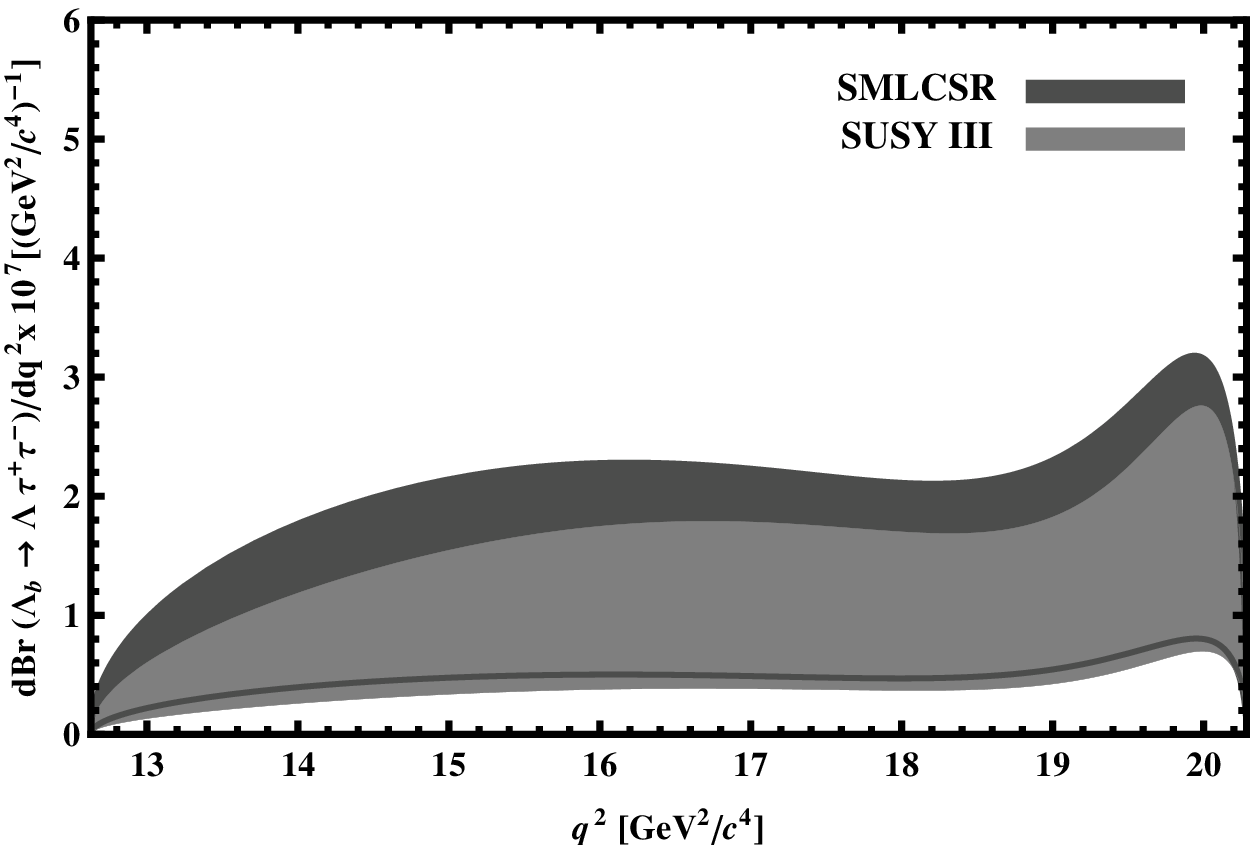,width=0.45\linewidth,clip=} &
\epsfig{file=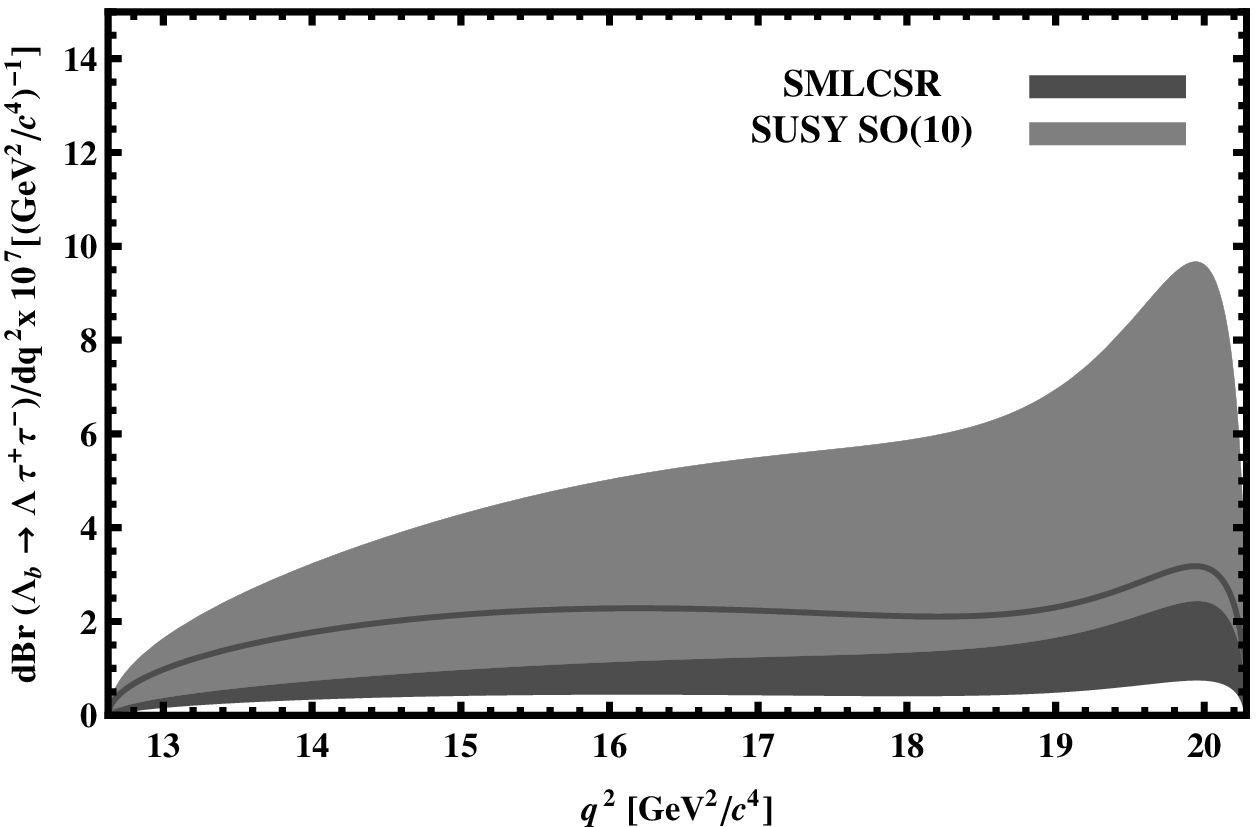,width=0.45\linewidth,clip=} 
\end{tabular}
\caption{The dependence of the differential branching ratio on  $q^2$  for the $\Lambda_{b}\rightarrow \Lambda \tau^{+} \tau^{-}$ 
transition in SMLCSR and  SUSY III and SO(10) models.}
\end{figure}

By using all these input parameters and the form factors with their uncertainties, we  present the dependence of the differential branching ratio of the $\Lambda_{b}\rightarrow \Lambda \ell^{+} \ell^{-}$  on  $q^2$ in SMLCSR, RS$_c$ and different SUSY models in figures 1-6.
In these figures we also show the experimental data provided by LHCb \cite{LHCb} as well as the existing lattice QCD predictions \cite{lattice}.
 We do not present the results 
for $e$ in the presentations since the predictions at $e$ channel are very close to those of $\mu$. 

 From figures 1-6 we see that 
\begin{itemize}
\item for all lepton channels, the SMLCSR and RS$_c$ models have roughly the same predictions except for some values of $q^2$ at which 
 there are small differences between predictions of the SMLCSR  and RS$_c$ models on the differential branching ratio.
  \item The areas swept by the SMLCSR  are wider compared to those of lattice QCD \cite{lattice} existing in the $\mu$ channel but they include
 those predictions.
\item The experimental data in the intervals $4$ GeV$^2/$c$^4$ $\leq q^2\leq 6$ GeV$^2/$c$^4$ and  $18$ GeV$^2/$c$^4$ $\leq q^2\leq 20$ GeV$^2/$c$^4$
cannot be described by the SMLCSR, lattice QCD or RS$_c$ models. In the remaining intervals the SMLCSR, lattice and RS$_c$ models reproduce
the experimental data, except for $6$ GeV$^2/$c$^4$ $\leq q^2\leq 8$ GeV$^2/$c$^4$, for which the datum remains outside of the lattice predictions.
 \item In the $\tau$ channel, the bands of the SMLCSR and RS$_c$ scenarios intersect each other, except for higher values of $q^2$, for which 
 the  errors of the form factors do not kill the differences between the two model predictions.
 \item At all lepton channels, the SUSY models show overall considerable deviations from the SMLCSR, lattice QCD and experimental data although they include the predictions of these
 models for some values of $q^2$. The maximum deviations of the SUSY predictions from the results of SMLCSR, lattice QCD and experiment belongs to the 
 SUSY II such that the SMLCSR, lattice QCD and experimental results remain out of the regions swept by the SUSY II model at higher values of $q^2$.
  \item  In the $\mu$ channel, the experimental  data   in the interval $18$ GeV$^2/$c$^4$ $\leq q^2\leq 20$ GeV$^2/$c$^4$ are reproduced by SUSY I, III and SO(10) but
  not by SUSY II. Note that in this interval other models (SMLCSR, lattice QCD and RS$_c$) also can not describe the experimental data.
 \item Again in the $\mu$ channel, the experimental data in the interval $4$ GeV$^2/$c$^4$ $\leq q^2\leq 6$ GeV$^2/$c$^4$ cannot be reproduced
 by any SUSY models like the SMLCSR, lattice QCD and RS$_c$ scenarios.
 \item In the case of $\tau$ as the final lepton, there are considerable differences
 between different SUSY models' predictions and that of the SMLCSR  and these cannot be completely killed by the errors of form factors.
 The maximum deviations of the SUSY results from the SMLCSR predictions belong to the SUSY II at higher $q^2$ values.
 \end{itemize}
                                 %%%%%%%%%%%%%%%%%%%%%%%%%%%%%%%%%%%%%%%
       %%%%%%%%%%%%%%%%%%%%%%%%%%%%%%%%%%%%%%%             %%%%%%%%%%%%%%%%%%%%%%%%%%%%%%%%%%%%%%%%%%
                                 %%%%%%%%%%%%%%%%%%%%%%%%%%%%%%%%%%%%%%%

\subsection{The lepton forward-backward asymmetry }
In this subsection, we would like to present the results of the lepton forward-backward asymmetry obtained in different scenarios.
The lepton ${\cal A}_{FB}$  is defined  as
\bea {\cal A}_{FB} (\hat s)=
\frac{\ds{\int_0^1\frac{d^{2}\Gamma}{d\hat{s}dz}}(z,\hat s)\,dz -
\ds{\int_{-1}^0\frac{d^{2}\Gamma}{d\hat{s}dz}}(z,\hat s)\,dz}
{\ds{\int_0^1\frac{d^{2}\Gamma}{d\hat{s}dz}}(z,\hat s)\,dz +
\ds{\int_{-1}^0\frac{d^{2}\Gamma}{d\hat{s}dz}}(z,\hat s)\,dz}~. 
\eea
\begin{figure}[h!]
\centering
\begin{tabular}{cc}
\epsfig{file=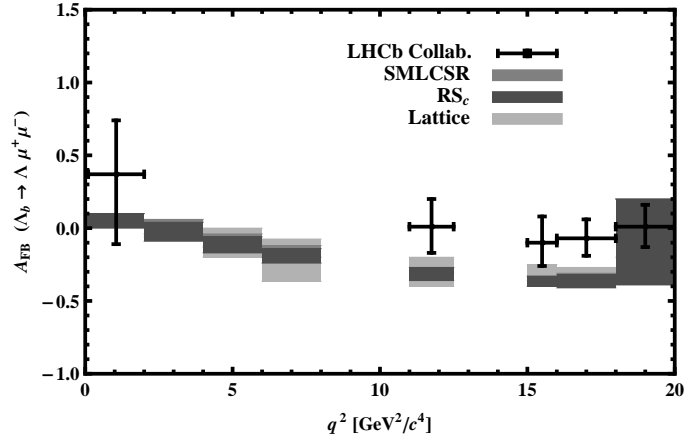,width=0.55\linewidth,clip=}  
\end{tabular}
\caption{The dependence of the ${\cal A}_{FB}$ on  $q^2$  for $\Lambda_{b}\rightarrow \Lambda \mu^{+} \mu^{-}$ transition 
in  SMLCSR, lattice QCD \cite{lattice} and RS$_c$ models together with 
recent experimental data by LHCb \cite{LHCb}.}
\end{figure}
\begin{figure}[h!]
\centering
\begin{tabular}{cc}
\epsfig{file=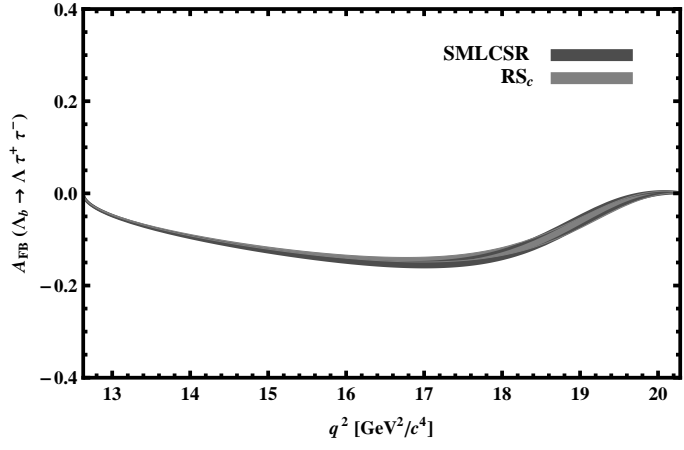,width=0.55\linewidth,clip=}  
\end{tabular}
\caption{The dependence of the ${\cal A}_{FB}$ on  $q^2$  for $\Lambda_{b}\rightarrow \Lambda \tau^{+} \tau^{-}$ transition 
in  SMLCSR and RS$_c$ models.}
\end{figure}
\begin{figure}[h!]
\centering
\begin{tabular}{ccc}
\epsfig{file=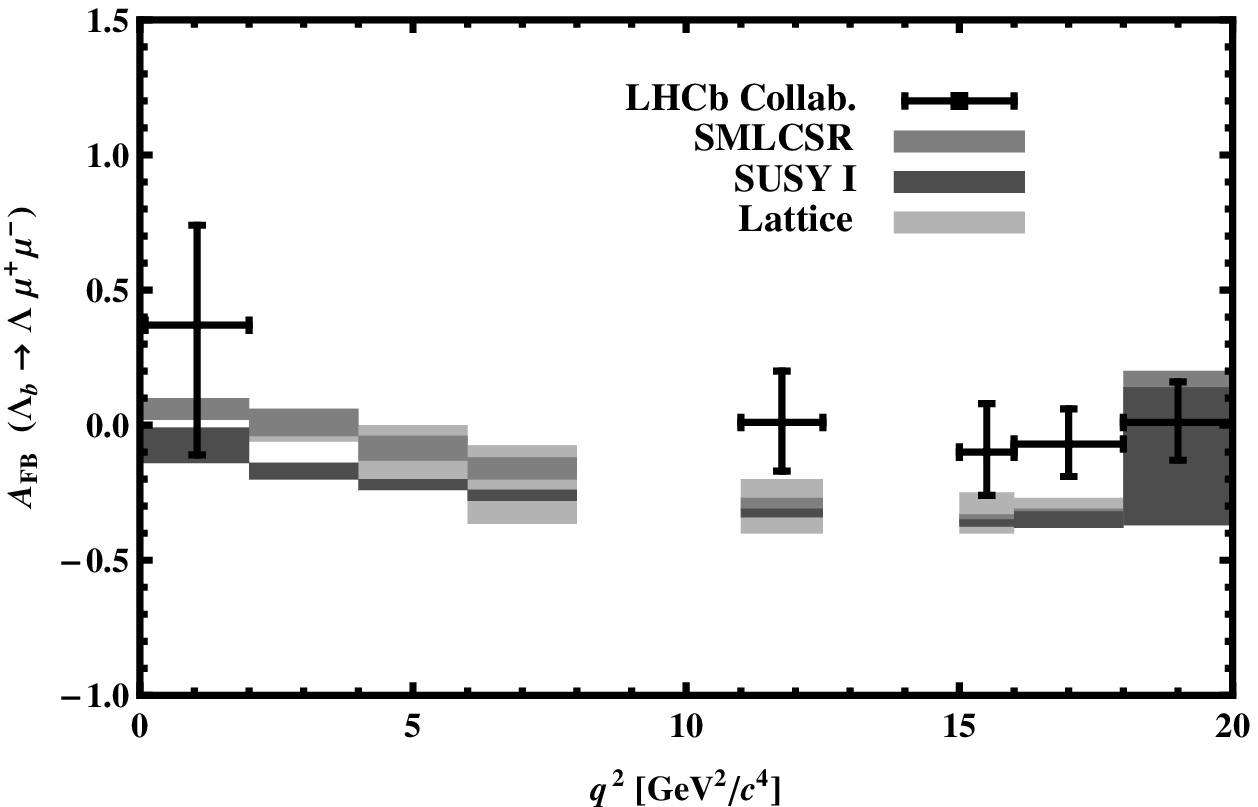,width=0.45\linewidth,clip=}&
\epsfig{file=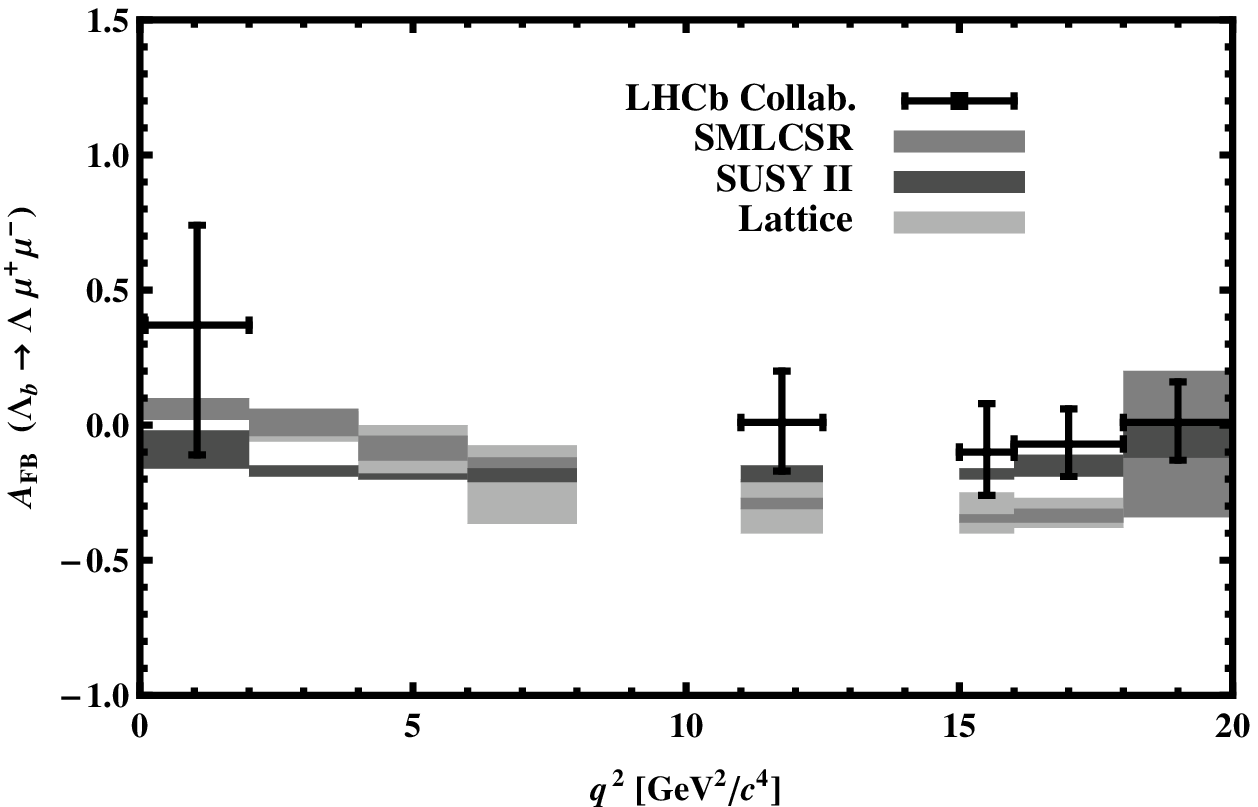,width=0.45\linewidth,clip=}
\end{tabular}
\caption{The dependence of the ${\cal A}_{FB}$ on  $q^2$  for $\Lambda_{b}\rightarrow \Lambda \mu^{+} \mu^{-}$ transition in SMLCSR,
 lattice QCD \cite{lattice} and SUSY I and  II  together with recent experimental data by LHCb \cite{LHCb}.}
\end{figure}

\begin{figure}[h!]
\centering
\begin{tabular}{ccc}
\epsfig{file=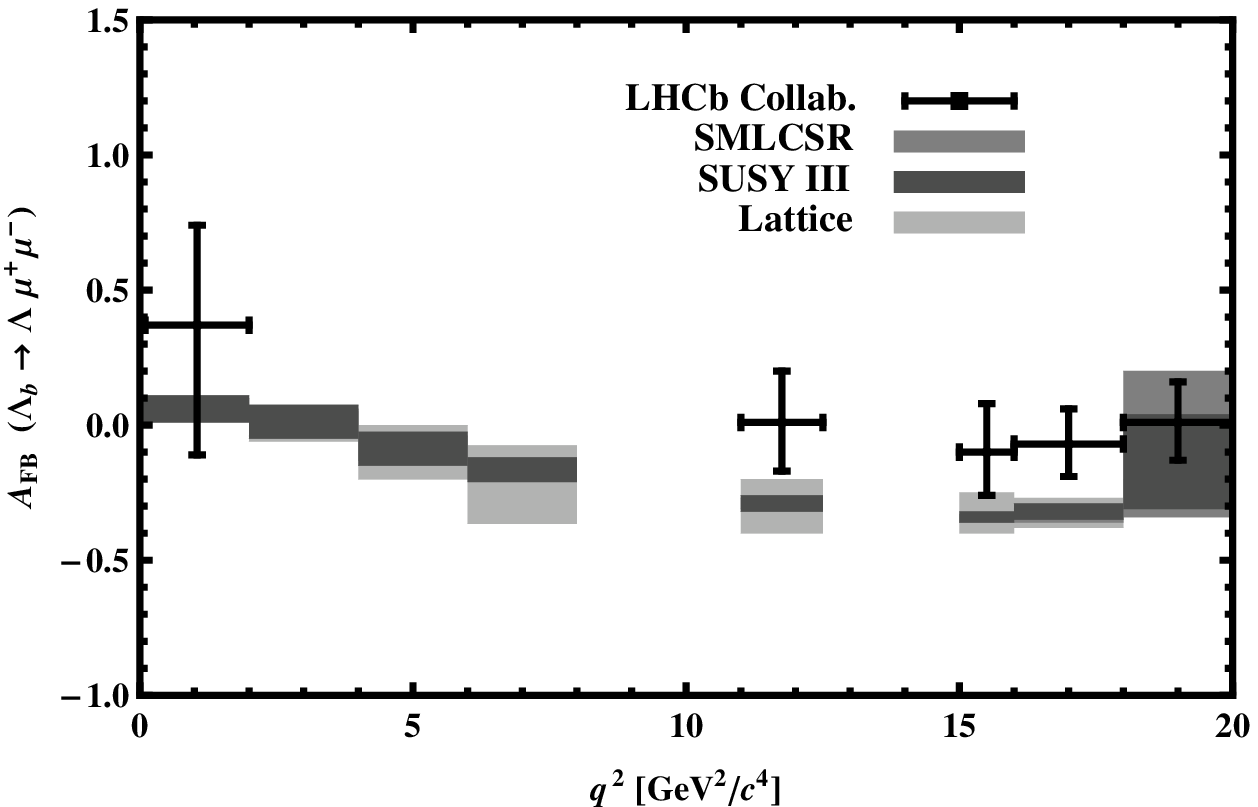,width=0.45\linewidth,clip=}&
\epsfig{file=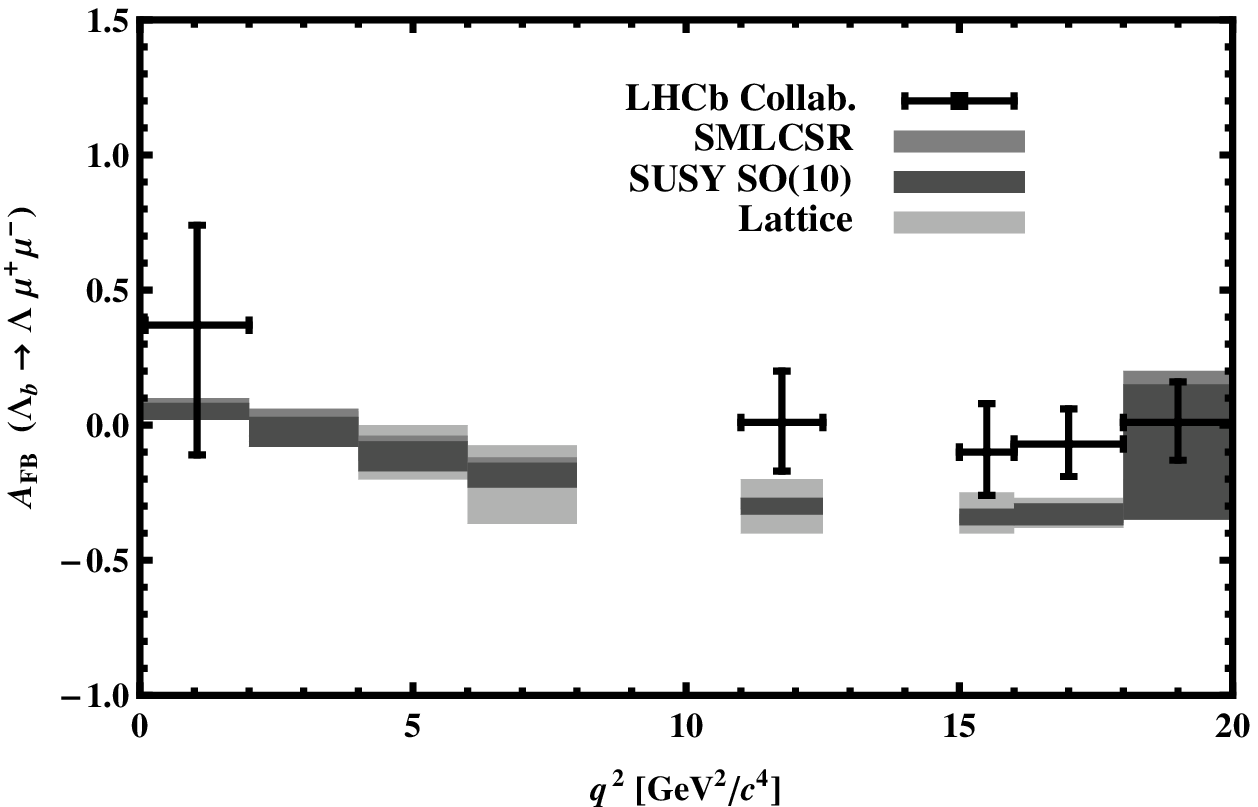,width=0.45\linewidth,clip=}
\end{tabular}
\caption{The dependence of the ${\cal A}_{FB}$ on  $q^2$  for $\Lambda_{b}\rightarrow \Lambda \mu^{+} \mu^{-}$ transition in SMLCSR,
 lattice QCD \cite{lattice} and SUSY III and SO(10)  together with recent experimental data by LHCb \cite{LHCb}.}
\end{figure}

\begin{figure}[h!]
\centering
\begin{tabular}{ccc}
\epsfig{file=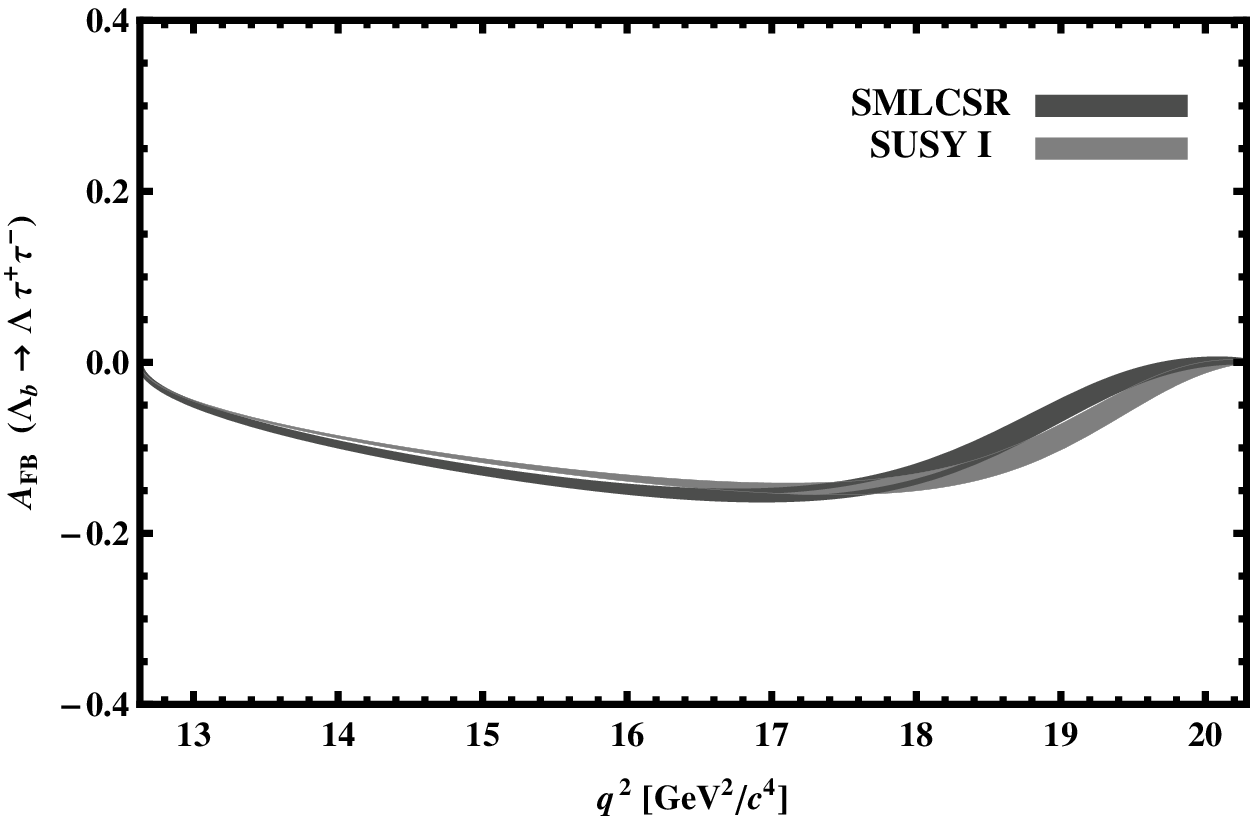,width=0.45\linewidth,clip=}&
\epsfig{file=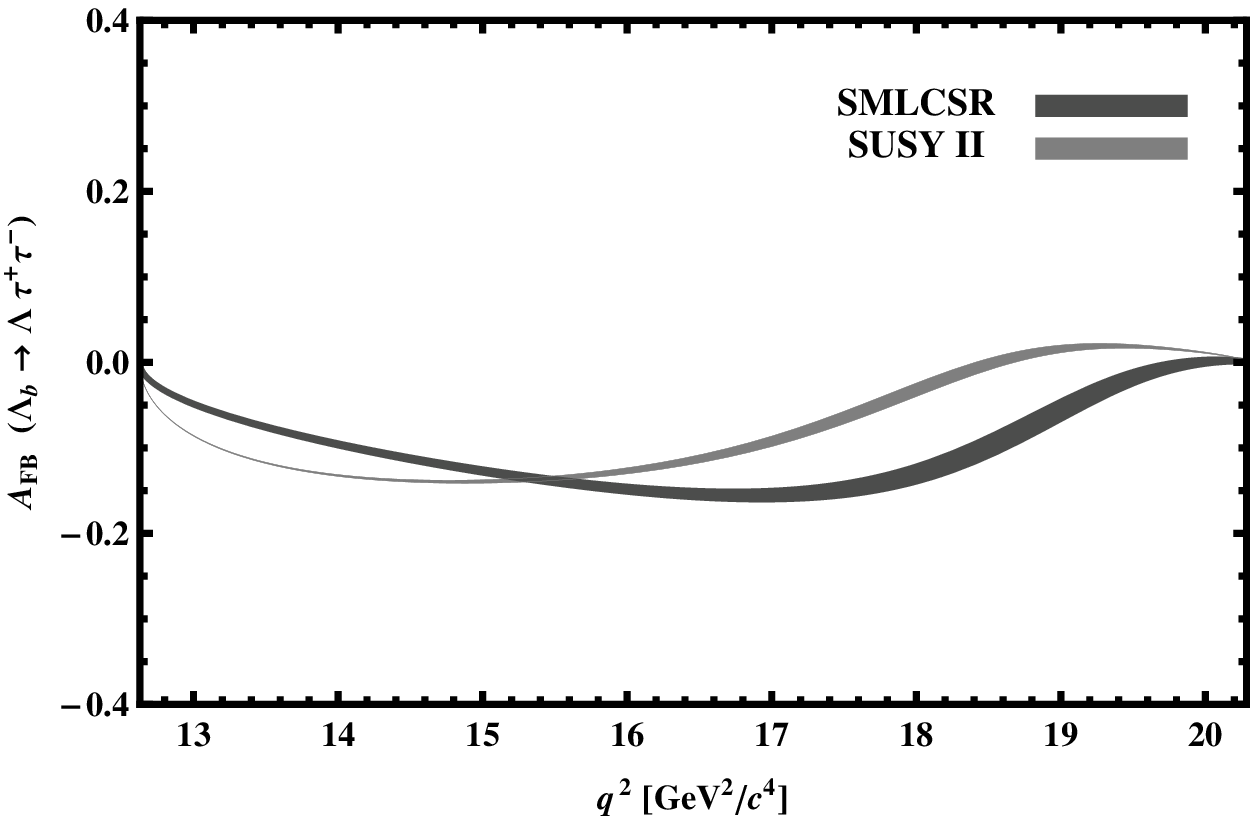,width=0.45\linewidth,clip=}
\end{tabular}
\caption{The dependence of the ${\cal A}_{FB}$ on  $q^2$  for $\Lambda_{b}\rightarrow \Lambda \tau^{+} \tau^{-}$ transition in SMLCSR and 
SUSY I and II scenarios.}
\end{figure}

\begin{figure}[h!]
\centering
\begin{tabular}{ccc}
\epsfig{file=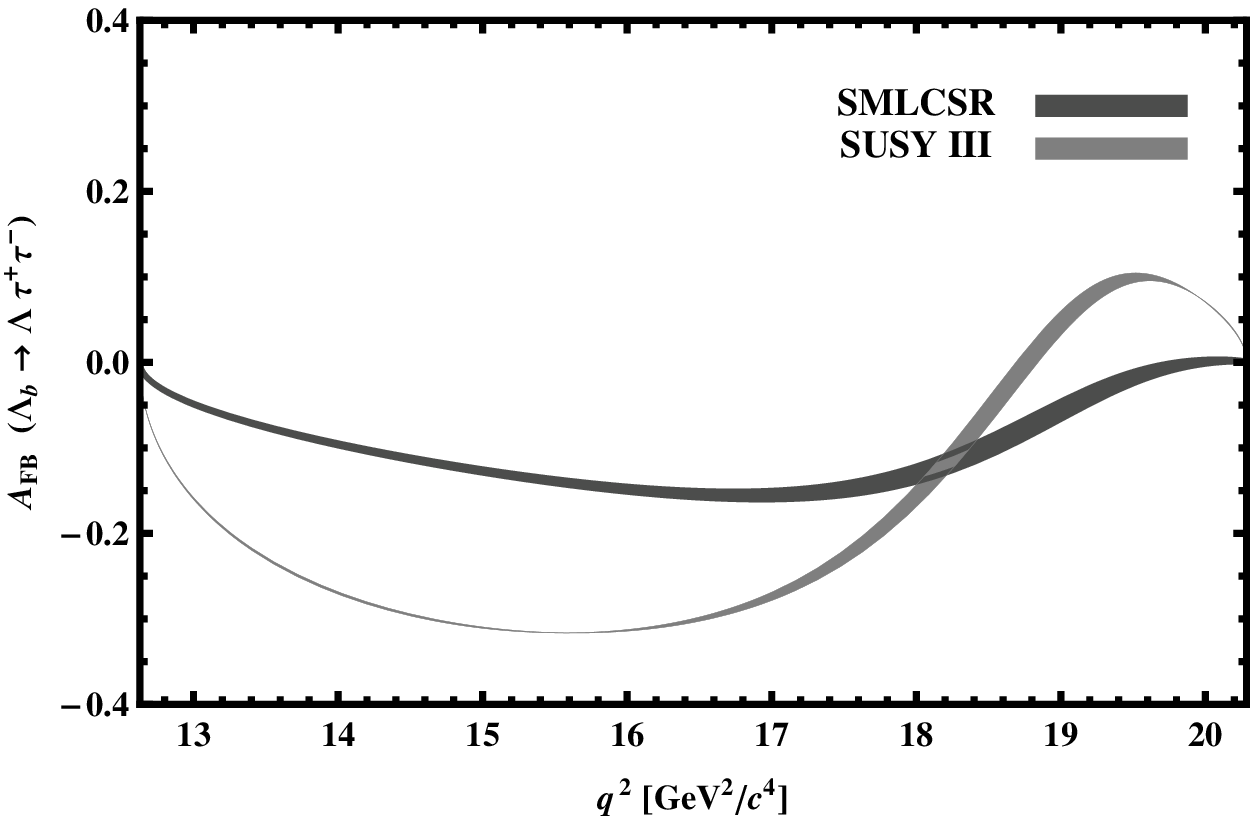,width=0.45\linewidth,clip=}&
\epsfig{file=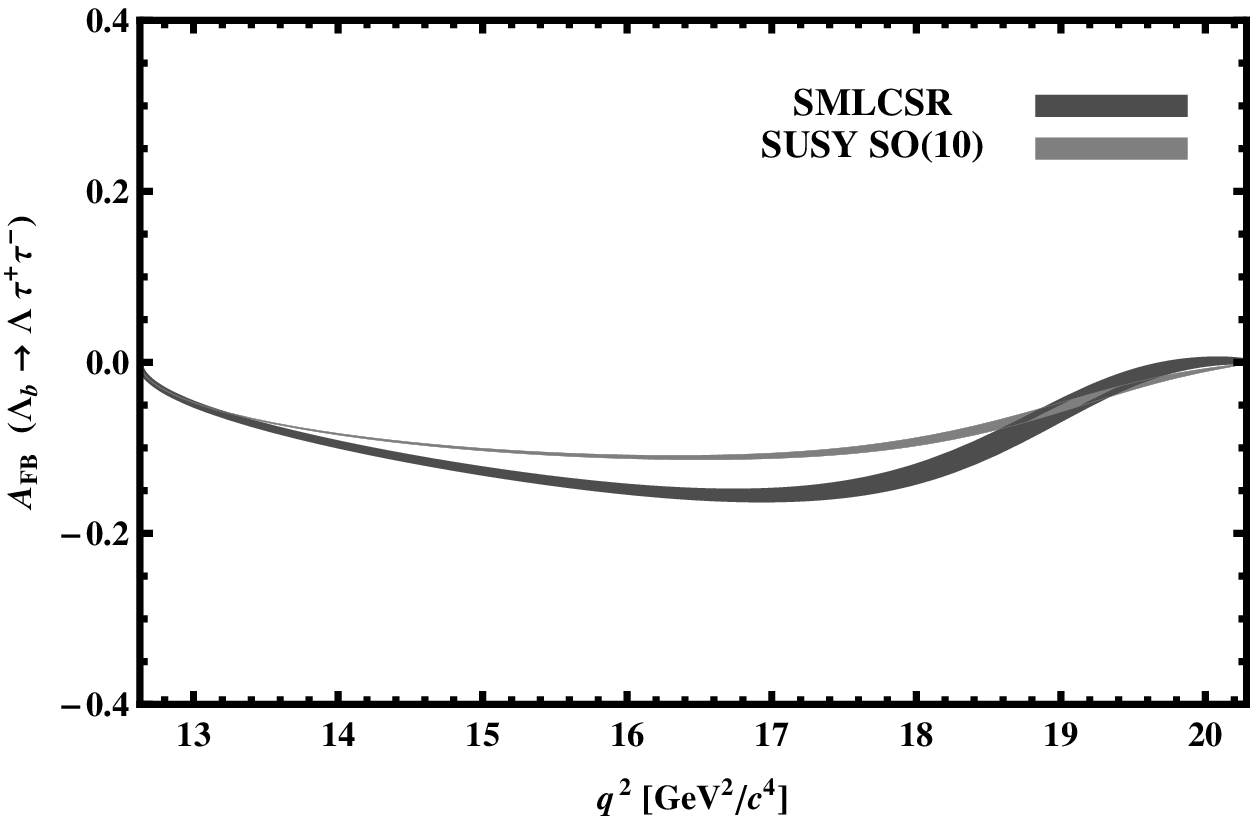,width=0.45\linewidth,clip=}
\end{tabular}
\caption{The dependence of the ${\cal A}_{FB}$ on  $q^2$  for $\Lambda_{b}\rightarrow \Lambda \tau^{+} \tau^{-}$ transition in SMLCSR and 
SUSY III and SO(10) scenarios.}
\end{figure}
Considering the form factors with their uncertainties from \cite{form-factors},  we  plot the dependence of the lepton forward-backward asymmetry on  $q^2$  for the decay under consideration
in both lepton channels in the SMLCSR, RS$_c$ and 
different SUSY models in figures 7-12.
 From these figures, we obtain that
\begin{itemize}
\item in the $\mu$ channel, the SMLCSR, lattice QCD and RS$_c$ models predictions on ${\cal A}_{FB}$ coincide with each other.
Except for the lattice QCD, they can only describe the experimental data existing in the $0$ GeV$^2/$c$^4$ $\leq q^2\leq 2$ GeV$^2/$c$^4$ and  $18$ GeV$^2/$c$^4$ $\leq q^2\leq 20$ GeV$^2/$c$^4$ regions. The remaining
data lie out of the swept regions by all these models.
\item As far as the SUSY models are considered, in the $\mu$ channel, the SUSY models have predictions that deviate from the SMLCSR and lattice
QCD predictions, considerably. All SUSY models reproduce the experimental data in the regions $0$ GeV$^2/$c$^4$ $\leq q^2\leq 2$ GeV$^2/$c$^4$ and  $18$ GeV$^2/$c$^4$ $\leq q^2\leq 20$ GeV$^2/$c$^4$.
The other data also remain out of the regions swept by different SUSY models except for the SUSY II, which reproduces the experimental data also 
in the interval $15$ GeV$^2/$c$^4$ $\leq q^2\leq 18$ GeV$^2/$c$^4$.
\item  In the case of the $\tau$ lepton, the SMLCSR and RS$_c$ have roughly the same predictions on  ${\cal A}_{FB}$.
\item In the $\tau$ lepton channel, the SMLCSR and SUSY I have roughly the same predictions for  ${\cal A}_{FB}$; however, the remaining
SUSY models' predictions deviate from the SMLCSR predictions considerably, although they intersect each other at some points.
\end{itemize}

                            %%%%%%%%%%%%%%%%%%%%%%%%%%%%%%%%%%%%%%%
       %%%%%%%%%%%%%%%%%%%%%%%%%%%%%%%%%%%%%%%             %%%%%%%%%%%%%%%%%%%%%%%%%%%%%%%%%%%%%%%%%%
                                 %%%%%%%%%%%%%%%%%%%%%%%%%%%%%%%%%%%%%%%
\section{Conclusion}
In the present work, we have analyzed the semileptonic $\Lambda _b\rightarrow \Lambda \ell^+ \ell^-$  decay mode  in SMLCSR, different SUSY models
and the RS$_c$ scenario. Using the form factors calculated in light cone QCD sum rules in the full theory \cite{form-factors}, 
we evaluated the differential branching ratio   and lepton forward-backward asymmetry 
for different leptons in those scenarios.
We also compared the results obtained via SMLCSR, RS$_c$ and different SUSY scenarios with the recent experimental data provided by LHCb
\cite{LHCb}  as well as the existing lattice QCD predictions \cite{lattice} 
on the considered quantities. 
We observed that the regions swept by the SMLCSR model include the RS$_c$ predictions although they are somewhat wider compared to those of  RS$_c$ models for the 
considered physical quantities. The SMLCSR predictions on the considered quantities in the present work are overall consistent with the lattice
QCD predictions provided by Ref. \cite{lattice}. 

The predictions of different SUSY models on the differential branching ratio deviate considerably from the SMLCSR and lattice predictions. The maximum
deviations belong to the SUSY II model. In the case of ${\cal A}_{FB}$ and the $\mu$ channel,  the predictions of different SUSY models have considerable
deviations from the SMLCSR and lattice QCD predictions. For ${\cal A}_{FB}$ and the $\tau$ channel, the SUSY I and SMLCSR have roughly the same predictions but 
the other SUSY models have predictions different from that of the SMLCSR.

 The experimental data on the differential branching ratio in the $\mu$ channel can be reproduced
by SMLCSR, lattice QCD and RS$_c$ models except for the intervals $4$ GeV$^2/$c$^4$ $\leq q^2\leq 6$ GeV$^2/$c$^4$ 
and $18$ GeV$^2/$c$^4$ $\leq q^2\leq 20$ GeV$^2/$c$^4$, which cannot be described by SMLCSR, lattice QCD or RS$_c$ models. As far as
the SUSY models are considered, different SUSY models also cannot reproduce the experimental data in the interval $4$ GeV$^2/$c$^4$ $\leq q^2\leq 6$ GeV$^2/$c$^4$.
However, except for SUSY II, the remaining SUSY scenarios can explain the experimental data in the region $18$ GeV$^2/$c$^4$ $\leq q^2\leq 20$ GeV$^2/$c$^4$.

In the case of  ${\cal A}_{FB}$ and the $\mu$ channel, the SMLCSR, RS$_c$ and different SUSY models can only describe the experimental data existing in the $0$ GeV$^2/$c$^4$ $\leq q^2\leq 2$ GeV$^2/$c$^4$ and 
$18$ GeV$^2/$c$^4$ $\leq q^2\leq 20$ GeV$^2/$c$^4$ regions. The other existing data remain out of the swept areas by these models, except for SUSY II, which  can also 
reproduce the experimental data in $15$ GeV$^2/$c$^4$ $\leq q^2\leq 18$ GeV$^2/$c$^4$. 

More experimental data 
in the $\mu$ channel related to different physical quantities associated with the $\Lambda _b\rightarrow \Lambda \mu^+ \mu^-$ mode,
the future experimental data in the $\tau$ channel and comparison of the results with our predictions on the quantities considered in the present work may help us in
the course of searching for new physics effects. 
                                 %%%%%%%%%%%%%%%%%%%%%%%%%%%%%%%%%%%%%%%
       %%%%%%%%%%%%%%%%%%%%%%%%%%%%%%%%%%%%%%%             %%%%%%%%%%%%%%%%%%%%%%%%%%%%%%%%%%%%%%%%%%
                                 %%%%%%%%%%%%%%%%%%%%%%%%%%%%%%%%%%%%%%%
%\section{Acknowledgement}
                                 
                                 %%%%%%%%%%%%%%%%%%%%%%%%%%%%%%%%%%%%%%%
       %%%%%%%%%%%%%%%%%%%%%%%%%%%%%%%%%%%%%%%             %%%%%%%%%%%%%%%%%%%%%%%%%%%%%%%%%%%%%%%%%%
                                 %%%%%%%%%%%%%%%%%%%%%%%%%%%%%%%%%%%%%%%
                            
                                 %%%%%%%%%%%%%%%%%%%%%%%%%%%%%%%%%%%%%%%
       %%%%%%%%%%%%%%%%%%%%%%%%%%%%%%%%%%%%%%%             %%%%%%%%%%%%%%%%%%%%%%%%%%%%%%%%%%%%%%%%%%
                                 %%%%%%%%%%%%%%%%%%%%%%%%%%%%%%%%%%%%%%%
\end{document}